\documentclass[a4paper,11pt]{article}
\pdfoutput=1 

\usepackage{jheppub1} 

\usepackage[T1]{fontenc} 

\usepackage{amsmath,amssymb,amsfonts,graphicx,subfigure,bm}

\def\bea{\begin{eqnarray}}
\def\eea{\end{eqnarray}}
\def\nn{\nonumber}
\def\ba{\begin{array}}
\def\ea{\end{array}}
\def\nn{\nonumber}
\def\Tr{\text{Tr}}

\def\sgn{\text{sgn}}

\def\Pf{\text{Pf}}

\def\sn{\text{sn}}
\def\dn{\text{dn}}

\title{\boldmath Phase transition in von Neumann entanglement entropy from replica symmetry breaking}

\author[]{Shao-Kai Jian, }
\author[]{Brian Swingle}

\affiliation[]{Department of Physics, Brandeis University, Waltham, Massachusetts 02453, USA}
\affiliation[]{Condensed Matter Theory Center and Joint Quantum Institute, Department of Physics, University of Maryland, College Park, Maryland 20742, USA}

\emailAdd{skjian@brandeis.edu, bswingle@umd.edu}

\abstract{We study the entanglement transition in monitored Brownian SYK chains in the large-$N$ limit. Without measurement the steady state $n$-th R\'enyi entropy is obtained by summing over a class of solutions, and is found to saturate to the Page value in the $n\rightarrow 1$ limit. In the presence of measurements, the analytical continuation $n\rightarrow 1$ is performed using the cyclic symmetric solution. The result shows that as the monitoring rate increases, a continuous von Neumann entanglement entropy transition from volume-law to area-law occurs at the point of replica symmetry unbreaking.}

\begin{document} 
\maketitle
\flushbottom

\section{Introduction}
Starting from an unentangled product state, a chaotic Hamiltonian or circuit generates entanglement between different parts of the system, and eventually leads to a state with volume-law entanglement at late time. On the contrary, local measurements extract information and diminish the entanglement between the measured qubits and the rest of the system. The competition between these two effects leads to a transition in the degree of entanglement in the long-time steady state, the so-called measurement-induced phase transition~\cite{li2018quantum, li2019measurement, skinner2019measurement, gullans2020dynamical, chan2019unitary}.

Following this realization, there has been a surge of investigations of this transition and of related phenomena~\cite{vasseur2019entanglement,choi2020quantum,jian2020measurement,bao2020theory,zabalo2020critical,gullans2020scalable,li2020conformal,iaconis2020measurement,nahum2021measurement,li2021statistical,sang2021measurement,lavasani2021measurement,ippoliti2021entanglement,bao2021symmetry,bentsen2021measurement,jian2021syk,li2021entanglement,jian2021quantum,yang2021entanglement}. In principle, one can diagnose transitions in entanglement properties by introducing replicas and using them to compute the $n$-th R\'enyi entropy of the long-time steady state. In this context, it was found that the entanglement transition can be understood as an unbreaking transition of replica symmetry~\cite{vasseur2019entanglement,jian2020measurement,bao2020theory,nahum2021measurement}. Nevertheless, as the von Neumann entanglement entropy is obtained in the $n\rightarrow 1$ limit, where $n$ is the number of replicas, how the replica symmetry unbreaking defined for $n\ge 2$ interplays with the replica limit $n \rightarrow 1$ is still an outstanding question. Part of the difficulty lies in the hardness of analytical continuation. In the context of Haar random hybrid circuits, the transition can be effectively mapped to a statistical model in the replica space, which in the limit of infinite local Hilbert space dimension is described by bond percolation~\cite{skinner2019measurement,jian2020measurement,bao2020theory}, and the replica limit $n\rightarrow 1$ is notoriously subtle even in this limit. In this paper, we fill this gap by introducing a monitored Brownian SYK model~\cite{kitaev2015simple,sachdev1993gapless,maldacena2016remarks,saad2018semiclassical,sunderhauf2019quantum,liu2021non,jian2020note,jian2021syk,stanford2021subleading}, and calculating the von Neumann entanglement entropy using the replica trick. 

This model, which features continuous weak monitoring, admits a path integral representation, and consequently allows saddle-point analysis in the large-$N$ limit. We show explicitly how summing over a class of replica symmetry broken solutions can correctly produce the Page value~\cite{page1993average} of the entanglement entropy in the absence of monitoring. A crucial step in the calculation is a mapping of part of the saddle point evaluation to the evaluation of a transition amplitude of a quantum mechanical model. Turning on measurement decreases the degree of replica symmetry breaking and eventually leads to its unbreaking. We then take the $n\rightarrow 1$ limit based on the cyclic symmetric solution and extract the behavior of the entanglement entropy per site as a function of measurement strength. The transition in the von Neumann entropy density in the $n\rightarrow 1$ limit coincides with the unbreaking transition of replica symmetry for $n\ge 2$. 

The paper is organized as follows. In Section~\ref{sec:quasi}, we define the quasi entropy of a monitored system~\cite{napp2019efficient}. Because of the non-unitary nature of measurements, a proper density matrix needs to be normalized after each measurement. This leads to difficulties in analytical calculations as the normalization is a nonlinear procedure that depends on the measurement outcome. The quasi entropy is the R\'enyi entropy of the unnormalized density matrix weighted such that it vanishes for a pure state.  The quasi entropy has two crucial properties: it admits a path integral representation and, like the R\'enyi entropy, the $n \rightarrow 1$ limit gives the von Neumann entanglement entropy averaged over quantum trajectories. 

In Section~\ref{sec:model}, we introduce our model which contains both a chaotic unitary part and a measurement part. The unitary part is given by two independent Brownian SYK chains. The Brownian SYK Hamiltonian is able to efficiently generate entanglement among different parts of the chain. The measurement part amounts to weakly projecting onto the Fermi parity of the complex fermions formed by the left-right pair of Majoranas at each site. Importantly, it is a local measurement that extracts information about a single site of the left-right coupled system, and its dark state corresponds to a product state between different sites with definite Fermi parity. Therefore, the competition between the two parts will result in a measurement-induced entanglement transition.

In Section~\ref{sec:cluster}, we first study a simpler version of our model that only consists of the unitary part. It is given by two coupled Brownian SYK clusters. After properly defining the initial state, the entanglement entropy between the two clusters can be obtained via a large-$N$ path integral in the replica space. We find a class of saddle points that are described by permutation matrices. The calculation of the onshell action can be done by decomposing the permutation into cycles, and then mapping each cycle to a transition amplitude of a Kitaev chain~\cite{kitaev2001unpaired}. An interesting point arising in the calculation is that the Pfaffian effectively counts the number of cycles in the solution given by a general permutation matrix. 
Then, by summing over this class of saddle points, we get a result that predicts maximal entanglement at leading order with subleading corrections consistent with the Page value~\cite{page1993average} and the symmetry of the model. 

In Section~\ref{sec:monitor}, we generalize the calculation into two Brownian SYK chains with a measurement that couples the two chains. The aforementioned class of saddle points that saturate the Page value are changed in the presence of the measurement. When the measurement rate gets larger than a critical value, all saddle points reduce to a diagonal solution where no correlation between different replica exists, i.e., the replica symmetry is restored. The calculation of quasi entropy now involves two coupled Kitaev chains, where the hopping within each chain is given by the correlation between different replicas, and the coupling between two chains is proportional to the measurement rate. After the coupling increases to a critical point, the hopping within each chain vanishes, and the eigenmodes become local. 
We take an analytical continuation of the cyclic symmetric solution to get the von Neumann entanglement entropy. The entropy density is obtained, and it shows a continuous phase transition right at the point where the replica symmetry is restored. 

We conclude the paper in Section~\ref{sec:conclusion}, where we also study the Landau-Ginzburg theory. We leave the technical details of the permutation operator that is used to calculate the quasi entropy, the explicit saddle-point solution for $n=2$, and the product of special trigonometric functions that is used to evaluate the transition amplitude to the appendices.

\section{Quasi entropy and trajectory averaged entanglement entropy} \label{sec:quasi}

We consider time evolution generated by unitary evolution and measurements, and are interested in the entanglement entropy of the long-time steady state. The state itself is not steady in a microscopic sense, but its entanglement structure is expected to stabilize at long times. Importantly, to really assess the entanglement properties of the resulting state, one needs to keep track of each measurement record. A specific evolution with given measurement outcomes forms a quantum trajectory. Ultimately, we will calculate the von Neumann entanglement entropy averaged over quantum trajectories.

As the unitary evolution part of the evolution is well known, here we focus on the measurement part. The measurement is described by a set of Kraus operator $\{ K_\nu \}$, $ \sum_\nu K_\nu^\dag K_\nu = 1$~\cite{wiseman1996quantum,jian2020measurement}, where $\nu$ numerates possible measurement outcomes. 
Starting from an initial density matrix $\rho$, given a sequence of measurement outcomes $\bm{\nu} = \nu_1 \nu_2...\nu_m$, the corresponding quantum trajectory is 
\bea
    \tilde \rho_{\bm{\nu}} = K_{\nu_m} ... K_{\nu_2} K_{\nu_1} \rho K_{\nu_1}^\dag K_{\nu_2}^\dag ...  K_{\nu_m}^\dag. 
\eea
Note that $\tilde \rho_{\bm{\nu}}$ is an unnormalized density matrix, whose trace gives the probability of the measurement ourcome $\bm{\nu}$.
Summing the probabilities of all possible measurement outcomes gives unity, i.e., $\sum_{\bm{\nu}} \Tr(\tilde \rho_{\bm{\nu}}) = 1$. 
We are interested in calculating the quasi entropy of bipartite system $A \bar A$~\cite{napp2019efficient},
\bea \label{eq:quasi}
    S_A^{(n)} &=& \frac1{1-n} \log \frac{\sum_{\bm{\nu}}\Tr[(\tilde \rho_{\bm{\nu}})^{\otimes n} M_\text{cyc}(A)]}{\sum_{\bm{\nu}} (\Tr[\tilde \rho_{\bm{\nu}}])^n },
\eea
where $M_{\text{cyc}}(A)$ is the cyclic permutation operator acting on the subsystem $A$ of the $n$ replicas. Note that the summation over quantum trajectories appears both in the numerator and denominator. 

Although the quasi entropy is distinct from the R\'enyi entropy of the normalized state with measurement outcome $\bm{\nu}$, the limit $n \rightarrow 1$ of the quasi entropy converges to the averaged von Neumann entanglement entropy of the normalized state. To show this, recall that the R\'enyi entropy of subsystem $A$ for a given quantum trajectory $\bm{\nu}$ is
\bea
    S_A^{(n)}(\bm{\nu}) = \frac1{1-n} \log \frac{\Tr[(\tilde \rho_{\bm{\nu}})^{\otimes n} M_\text{cyc}(A)]}{(\Tr[\tilde \rho_{\bm{\nu}}])^n }.
\eea
The von Neumann entanglement entropy can be obtained by analytically continuing the R\'enyi entropy to $n \rightarrow 1$, i.e., $S_A(\bm{\nu}) = \lim_{n \rightarrow 1}S_A^{(n)}(\bm{\nu}) $. We have the following relation
\bea \label{eq:expand}
    \frac{\Tr[(\tilde \rho_{\bm{\nu}})^{\otimes n} M_\text{cyc}(A)]}{(\Tr[\tilde \rho_{\bm{\nu}}])^n } \approx 1+(1-n) S_A(\bm{\nu}) + \mathcal O\left((n-1)^2 \right).
\eea
Using this relation, we can take $n \rightarrow 1$ limit of the quasi entropy defined in~(\ref{eq:quasi}),
\bea
   \lim_{n \rightarrow 1} S_A^{(n)} &= & \lim_{n \rightarrow 1} \frac1{1-n} \log \frac{\sum_{\bm{\nu}}(1+(1-n) S_A(\bm{\nu})) (\Tr[\tilde \rho_{\bm{\nu}}])^n}{\sum_{\bm{\nu}} (\Tr[\tilde \rho_{\bm{\nu}}])^n } \\
    &=& \lim_{n \rightarrow 1}\frac1{1-n} \log \sum_{\bm{\nu}} \left( 1 + (1-n) S_A(\bm{\nu}) \right) \Tr[ \tilde \rho_{\bm{\nu}}] \\
    &=& \lim_{n \rightarrow 1} \frac1{1-n} \log \left( 1+ (1-n) \sum_{\bm{\nu}} \Tr[ \tilde \rho_{\bm{\nu}}]  S_A(\bm{\nu}) \right) \\
    &=& \sum_{\bm{\nu}} \Tr[ \tilde \rho_{\bm{\nu}}] S_A(\bm{\nu}),
\eea
where in the first step we use~(\ref{eq:expand}) and in the second step and third step we use $\sum_{\bm{\nu}} \Tr[ \tilde \rho_{\bm{\nu}}] = 1$. Therefore, the $n\rightarrow 1$ limit of the quasi entropy is the quantum trajectory averaged von Neumann entanglement entropy. In the following, we will calculate the quasi entropy for arbitrary $n$, and then take the $n \rightarrow 1$ limit to get the von Neumann entanglement entropy.

\section{Model and setup} \label{sec:model} 

Our model consists of two coupled Brownian SYK chains, with a unitary part and non-unitary monitoring part. The unitary evolution is governed by the following Brownian SYK Hamiltonian describing left ($L$) and right ($R$) chains,
\bea \label{eq:LRhamiltonian}
	H(\psi) &=& \sum_{x;a=L,R}  \Big(  \sum_{i<j}i J_{a,ij}^{x}(t) \psi_{x,a,i} \psi_{x,a,j} \nn \\
	&& + \sum_{j_1,...,j_q} i^{q/2} U_{a,j_1...j_q}^{x,x+1}(t) \psi_{x,a,j_1}...\psi_{x,a,j_{q/2}} \psi_{x+1,a,j_{q/2+1}}... \psi_{x+1,a,j_q}  \Big),
\eea
where $\psi_{x,a,i}$ $i=1,...,N$ denotes $i$-th of $N$ Majorana fermions at each site $x=1,...,L$ of the $a = L, R$ chains, $\{ \psi_{x,a,j}, \psi_{x',a',j'}\} = \delta_{xx'} \delta_{aa'} \delta_{jj'}$. 
$L$ is the number of sites (it should not be confused with the $L$ chain), and periodic boundary conditions in real space are assumed in this paper. 
$J_{a,ij}^x$ is two Majorana coupling within each site, and $U^x_{a,j_1,...,j_q}$ is $q$ Majorana interaction between nearest-neighbor sites. Here we consider $q = 4 k, k \in \mathbb Z^+$ ($\mathbb Z^+ $ denotes positive integers) to conserve local Fermi parity. The couplings in the left and right chains are {\it independent} Gaussian variables with mean zero and variances
\bea \label{eq:random}
    \overline{J_{a,ij}^{x}(t_1) J_{a',ij}^{x'}(t_2)} &=& \frac{4J}{N}  \delta(t_{12})  \delta_{aa'} \delta^{x,x'}, \\
    \overline{U_{a,j_1...j_q}^{x,x+1}(t_1) U_{a',j_1...j_q}^{x',x'+1}(t_2)} &=& \frac{2^{q} (q/2)!^2 U}{q N^{q-1}} \delta(t_{12})  \delta_{aa'} \delta^{x,x'}. 
\eea
where $t_{12} = t_1 - t_2$. The Dirac $\delta$ function $\delta(t_{12})$ indicates the Brownian nature of the couplings. In the unitary evolution, the left and right chains do not couple, but the evolution does scramble information over the entire individual chains~\cite{jian2021quantum}.

To describe the monitoring part, we first divide the continuous time into infinitesimal steps and define proper Kraus operators, then we take the continuum limit to get an effective action description. In each infinitesimal time step $\delta t$, we consider the following measurement. The local measurement operator couples the $L$ annd $R$ fermions at each site, as described by the Kraus operators~\cite{jian2021syk}
\bea \label{eq:kraus}
\{ K_1^{x,i}, K^{x,i}_2\}= \left\{   \pi^-_{x,i} + \sqrt{1-\mathfrak{s}^2} \pi^+_{x,i}  ,\mathfrak{s} \pi^+_{x,i} \right\},
\eea
where $\pi^\pm_{x,i} = \frac{1}{2} (1 \mp i 2\psi_{x,L,i} \psi_{x,R,i})$ is the projection onto one of the Fermi parity eigenstates and $0 \le \mathfrak{s} \le 1$ is the measurement strength. We will see in the following that the appropriate continuous monitoring strength should be $\mathfrak{s} \propto \sqrt{\delta t}$. The Kraus operators satisfy the nomralization condition $ \sum_{j=1}^2 (K_j^{x,i})^\dag K^{x,i}_j = 1 $. 

During a single time step in $\delta t$, the evolution of the unnormalized density matrix for a quantum trajectory with measurement outcome $\nu_{x,i}$ for each flavor at each site is
\bea
    \tilde \rho_{\bm{\nu}}  = (\otimes_{x,i} K_{\nu_{x,i}}) \rho (\otimes_{x,i} K_{\nu_{x,i}}^\dag),
\eea
where on the right-hand side we suppress the superscript for the flavor and site in Kraus operators $K_j^{x,i}$ as they can be inferred from the measurement outcome. The quasi entropy requires $n$ replicas, i.e.,
\bea
    \tilde \rho_{\bm{\nu}}^{\otimes n} = \otimes_{\alpha=1}^n (\otimes_{x,i} K_{\nu_{x,i}}^\alpha)  \rho^\alpha (\otimes_{x,i} K_{\nu_{x,i}}^{\alpha\dag}), 
\eea 
where we have introduced a new superscript Greek letter $\alpha =1,...,n$ to denote different replicas (it should not be confused with the original superscript of the Kraus operator that is used to denote site and flavor). 
We will use the superscript Greek letter as the replica index throughout the paper. 
According to the definition of quasi entropy, we can change the order of the trajectory average and the trace over the density matrix, and consider the trajectory average first, namely, $\sum_{\bm{\nu}}\Tr[(\tilde \rho_{\bm{\nu}})^{\otimes n} \cdots ] = \Tr[\sum_{\bm{\nu}}(\tilde \rho_{\bm{\nu}})^{\otimes n} \cdots ]$ for any $\cdots$ independent of $\bm{\nu}$. The effect of the monitoring operator on the replicas is given by the following operator,
\bea 
    \sum_{\bm{\nu}} \otimes_{\alpha=1}^n (\otimes_{x,i} K_{\nu_{x,i}}^\alpha) \otimes (\otimes_{x,i} K_{\nu_{x,i}}^{\alpha \dag}) 
    &=& \otimes_{x,i} \sum_{\nu_{x,i}=1,2} \otimes_{\alpha=1}^n \otimes_{s=\pm} K_{\nu_{x,i},s}^\alpha,
\eea
where we have introduced a Keldysh-like contour index $s=\pm$ to denote $K_{\nu_{x,i},+}^\alpha = K_{\nu_{x,i}}^\alpha$, and $K_{\nu_{x,i},-}^\alpha = K_{\nu_{x,i}}^{\alpha\dag}$. Finally, the average over measurement outcomes is~\cite{jian2021syk}
\bea
    \otimes_{x,i} \sum_{\nu_{x,i}=1,2} \otimes_{\alpha,s} K_{\nu_{x,i},s}^\alpha  
    &\approx& \otimes_{x,i} \left( 1 - \frac{\mathfrak{s}^2}2 \sum_{\alpha=1}^n \sum_{s=\pm} \pi^{+,\alpha}_{x,i,s} \right) \\
    &\approx&  \otimes_{x,i} \exp   \left( - \frac{\mathfrak{s}^2}2 \sum_{\alpha,s} \pi^{+,\alpha}_{x,i,s}\right) \\
    & = & \exp  \left( \frac{\delta t \mu}{2}  \sum_{x,i,\alpha,s} i \psi_{x,L,i,s}^\alpha \psi_{x,R,i,s}^\alpha \right),
\eea
in which we have used the relation $\pi^+_{x,a,j} + \pi^-_{x,a,j} = 1$ and assumed $ \mathfrak{s} \ll 1$ and kept orders up to $\mathcal{O}(\mathfrak{s}^2)$. In the last line, we introduce $ \mu = \frac{\mathfrak{s}^2}{\delta t} $ as the effective monitoring strength. When the continuum limit $\delta t \rightarrow 0 $ is taken, $\mu$ is kept fixed. This means that the appropriate measurement strength is $ \mathfrak{s} \propto \sqrt{\delta t}$. 
Constants are neglected because they will not affect the dynamics. 
After taking the continuum limit, we arrive at the following action description of the monitoring,
\bea \label{eq:monitor}
    -I_\mu = \int dt \frac{i \mu}2 \sum_{x,i,\alpha,s} \psi_{x,L,i,s}^\alpha(t) \psi_{x,R,i,s}^\alpha(t),
\eea
where the summation over time steps results in an integral in the continuum limit.

Since our model contains not only trajectory average but also disorder average due to the Brownian coupling, we generalize the quasi entropy to also include disorder average,
\bea \label{eq:quasi_disorder}
    S_A^{(n)} &=& \frac1{1-n} \log \frac{ \mathbb E \Tr[(\tilde \rho_{\bm{\nu}})^{\otimes n} M_\text{cyc}(A)]}{ \mathbb E (\Tr[\tilde \rho_{\bm{\nu}}])^n },
\eea
where $\mathbb E$ denotes both the trajectory average and the disorder average in~(\ref{eq:random}).

\section{Entanglement entropy in two coupled Brownian SYK clusters} \label{sec:cluster}

Before we consider the full calculation of entanglement entropy with monitoring, let us first consider a simpler case without measurement: two coupled Brownian SYK clusters, considered as a single chain with only two sites called $x=1,2$. Because the $R$ chain plays no role in this calculation, we will suppress the chain index in this section, so the relevant fermions are $\psi_{x,j} = \psi_{x,L,j}$. We will divide the system into two parts, with all fermions at site $x=1$ in $A$ and all fermions at site $x=2$ in the complement $\bar A$, and refer to $x=1$ as $x \in A$, or simply $A$, and $x=2$ as $x \in \bar A$ or simply $\bar A$.

\subsection{R\'enyi entropy of a time-evolved EPR state}

We consider the following setup: starting from the tensor product of thermofield double (TFD) states in each of the clusters ($x=1,2$), we compute the quasi entropy between the two clusters at late time~\cite{penington2019replica,gu2017spread,chen2020replica,jian2020note}. Because Brownian random interactions do not conserve energy, we simply consider an infinite temperature TFD state or, equivalently, a fermionic EPR state for each cluster. To describe such a state, we double the Hilbert space by introducing two copies of the fermions, the original fermions $\psi_{x,j}$ and copies $\chi_{x,j}$, for both clusters $x=1,2$. The initial density matrix is given by
\bea
    \rho = |\text{EPR} \rangle  \langle \text{EPR} |, \quad    (\psi_{x,j} + i \chi_{x,j} )| \text{EPR} \rangle =0, \quad \forall x=1,2, \quad \forall j=1,...,N.
\eea
The time evolution is generated by the sum of $\psi$ and $\chi$ Hamiltonians, $ H(\psi)$ and $ H(-i \chi)$. This choice implies that $H(\psi) | \text{EPR} \rangle = H(-i\chi) |\text{EPR} \rangle$. The time evolved density matrix becomes
\bea\label{eq:rho_t}
    \rho(T) = U(T) \rho U^\dag(T), \quad U(T) = \mathcal{T} e^{-i \int_0^{T/2} dt (H(\psi)+H(-i\chi))}, 
\eea
where $\mathcal T$ denotes time ordering. This joint $\psi$-$\chi$ evolution of the EPR state for time $T/2$ is equivalent to just evolving with $U_\psi(T) = \mathcal{T} e^{-i\int_0^{T} dt H(\psi)}$ or $U_\chi(T) = \mathcal{T} e^{-i\int_0^{T} dt H(-i\chi)}$ for time $T$ by virtue of the identity $H(\psi) | \text{EPR} \rangle = H(-i\chi) |\text{EPR} \rangle$.

Without measurement, the quasi entropy is same as annealed average of the R\'enyi entropy, 
\bea \label{eq:renyi}
    S_A^{(n)} = \frac1{1-n} \log  \mathbb E \Tr[\rho^{\otimes n} M_\text{cyc}(A)].
\eea 
For the EPR state, this quantity is related to the two-point function of cyclic permutation operators. Notice that the cyclic permutation operator factorizes, $M_\text{cyc}(A) = M_\text{cyc}^\psi(A) \otimes M_\text{cyc}^\chi(A)$ for the doubled Hilbert space, where each of them is given by (see Appendix~\ref{append:permutation})
\bea
    M^\psi_\text{cyc}(A) = \prod_{x \in A} \prod_{j=1}^N e^{\frac\pi2 \psi^1_{x,j} \psi^2_{x,j}} e^{\frac\pi2 \psi^3_{x,j} \psi^4_{x,j}} ... e^{\frac\pi2 \psi^{n-3}_{x,j} \psi^{n-2}_{x,j}} e^{\frac\pi2 \psi^{n-1}_{x,j} \psi^n_{x,j}},
\eea
where the superscript denotes replica index, and the same for the $\chi$ fermions. Then the trace can be cast into
\bea
    \Tr[\rho(T)^{\otimes n} M_\text{cyc}(A)] &=& \langle \text{EPR}|^{\otimes n} U^\dag_\chi(T)^{\otimes n} [M_\text{cyc}^\psi(A) \otimes M_\text{cyc}^\chi(A)] U_\chi(T)^{\otimes n} |\text{EPR} \rangle^{\otimes n} \\
    &=& \langle \text{EPR}|^{\otimes n} M_\text{cyc}^{\chi\dag}(A) U^\dag_\chi(T)^{\otimes n} M_\text{cyc}^\chi(A) U_\chi(T)^{\otimes n}   | \text{EPR} \rangle^{\otimes n} \\
    &=&  2^{-nN/2}\Tr[  M_\text{cyc}^{\chi\dag}(A) U^\dag_\chi(T)^{\otimes n}  M_\text{cyc}^{\chi}(A) U_\chi(T)^{\otimes n}],
\eea
where in the first line we use $\rho(T) = U_\chi(T) |\text{EPR} \rangle \langle  \text{EPR} | U_\chi^\dag(T) $, in the second line we note $ M^{\chi}_\text{cyc}(A)|\text{EPR}\rangle^{\otimes n} = M^{\psi\dag}_\text{cyc}(A)|\text{EPR}\rangle^{\otimes n}  $, and in the last line because the operators contain only $\psi$ fields, we can trace over the $\chi$ Hilbert space. As is seen from the last expression, the quasi entropy of the EPR state at time $T$ is given by the averaged two-point correlation function of the cyclic permutation operators. If $T=0$, it is easy to see the trace is one, consistent with the initial state being a pure state.

To get a path integral representation of the trace, it is useful to write it as
\bea
    \Tr[\rho(T)^{\otimes n} M_\text{cyc}(A)] &=& \langle \text{EPR}|^{\otimes n} M_\text{cyc}^{\chi\dag}(A) [U^\dag_\chi(T) \otimes U_\psi(T)]^{\otimes n} M_\text{cyc}^\chi(A)  | \text{EPR} \rangle^{\otimes n}.
\eea
If there were not permutation operators, then the trace would be one since these two time evolutions cancel for the EPR state. One then recognizes $U_\psi(T)$ and $U^\dag_\chi(T)$ as the forward and backward evolution in Keldysh contour language. Indeed, the boundary conditions for the fields in conventional Keldysh field theory are~\cite{altland2010condensed}
\bea \label{eq:bc}
    \psi_{x,j,+}^\alpha(0) = - \psi^\alpha_{x,j,-}(0), \quad \psi_{x,j,+}^\alpha (T) = \psi_{x,j,-}^\alpha(T) , \quad \forall j=1,...,N, \quad \forall \alpha =1,...,n,
\eea
where the minus sign at time $t=0$ is from the fermionic coherent state path integral. Recall that for an EPR state, 
\bea
    (\psi^\alpha_{x,j} + i \chi^\alpha_{x,j}) |\text{EPR} \rangle^{\otimes n} = 0, \quad  \langle \text{EPR} |^{\otimes n} (\psi^\alpha_{x,j} - i \chi^\alpha_{x,j}) = 0 , \quad \forall j=1,...,N, \quad \forall \alpha = 1,..., n, \nn\\ 
\eea
which suggests that by taking $\psi_{x,j}^\alpha \rightarrow \psi_{x,j,+}$ and $i \chi_{x,j}^\alpha \rightarrow \psi_{x,j,-}^\alpha$, we can write the trace as a $n$-replicated Keldysh field theory with a total of $2n$ contours. In the following, we will use notation like $\psi_+$, $\psi_-$ for Keldysh fields, while notation like $\psi$, $\chi$ for operators defined in EPR states.

The presence of the permutation operator then changes the boundary conditions for the fermions in subsystem $A$, for $x\in A$,
\bea
    \psi^\alpha_{x,j} M_\text{cyc}^\chi(A) | \text{EPR} \rangle^{\otimes n} &=&  M_\text{cyc}^\chi(A) (-i \chi^\alpha_{x,j}) | \text{EPR} \rangle^{\otimes n} \\
    &=& M_\text{cyc}^\chi(A) (-i \chi^\alpha_{x,j}) M_\text{cyc}^{\chi\dag}(A)  M_\text{cyc}^\chi(A) | \text{EPR} \rangle^{\otimes n} \\
    &=& \sum_\beta \sgn(\alpha-\beta) \delta^{\alpha+1,\beta} (-i \chi^\beta_{x,j}) M_\text{cyc}^\chi(A) | \text{EPR} \rangle^{\otimes n}.
\eea
In the last line, when $\alpha = n$, the symbol means $\delta^{n+1,\beta} = \delta^{1,\beta}$. The $n$-replicated Keldysh field theory is coupled through boundary conditions due to the permutation operator. We can redefine the fields in the backward contours to have the conventional boundary conditions, and so bring the effect of the twisted boundary conditions to the bulk action, i.e.,
\bea \label{eq:redefine}
     \sum_\beta \sgn(\alpha-\beta) \delta^{\alpha+1,\beta} \psi_{x,j,-}^\beta \rightarrow \psi_{x,j,-}^\alpha, \quad \forall x \in A, \quad \forall \alpha = 1,...,n, \quad\forall j=1,...N.
\eea

With this field redefinition, we can derive the path integral, $\mathbb E \Tr[\rho^{\otimes n} M_\text{cyc}(A)] = \int D G D \Sigma e^{-I}$, where the $G$-$\Sigma$ action reads~\cite{penington2019replica,jian2021syk, jian2020note}
\bea \label{eq:action}
	- \frac{I}{N} &=& \sum_{x=1,2} \Big[ \log \Pf [ \hat S \partial_t  - \hat \Sigma_x] + \int dt_1 dt_2  \big( - \frac{1}2  \Sigma_{x,ss'}^{\alpha\beta} G_{x,ss'}^{\alpha\beta} +\frac{J}{8} \delta(t_{12}) c_{ss'} (2G_{x,ss'}^{\alpha\beta}(t_1,t_2))^2 \big) \Big] \nn\\
	&& + \frac{U}{2q} \int dt_1 dt_2 \delta(t_{12}) c_{ss'}(2G_{1,ss'}^{\alpha\beta}(t_1,t_2))^{q/2} M^{\alpha\gamma}_s M^{\beta\delta}_{s'} (2G_{2,ss'}^{\gamma\delta}(t_1,t_2))^{q/2},
\eea
where $t_{12} \equiv t_1 - t_2$, and $s =\pm $ is introduced to denote the forward and backward evolution. $S_{++} = 1$, $S_{--}=-1$, $S_{+-} = S_{-+} = 0$, and $c_{++}=c_{--} = - 1$, $c_{+-} = c_{-+} =  1$ capture the structure of the forward and backward evolutions. The summation over the replica indices and the contour indices is implicit. The $M$ matrix is defined as $M^{\alpha\beta}_+ = \delta^{\alpha\beta}$, $M^{\alpha\beta}_- = \epsilon^{\alpha\beta}$, where $\epsilon^{\alpha\beta} \equiv \delta^{\alpha+1, \beta}$ is the cyclic permutation matrix originating from the redefinition in~(\ref{eq:redefine}). Because our model conserves the local Fermi parity, the sign from the redefinition~(\ref{eq:redefine}) disappears. The boundary conditions are the same as~(\ref{eq:bc}). In the large-$N$ action, there is an emergent $SO(N)$ symmetry among the flavors of Majorana $\psi_{x,j,s}^\alpha$, $j = 1,...,N$. For simplicity, we will often talk about the Majorana fermions simply as $\psi_{x,s}^\alpha$ without referring to each individual flavor.

\subsection{Saddle-point solutions given by permutations}

The saddle-point equations following from~(\ref{eq:action}) are
\bea
\label{eq:SD1}	\hat G^{-1}_x &=& \hat S \partial_t - \hat \Sigma_x, \\
\label{eq:SD2}	\Sigma_{x,ss'}^{\alpha\beta} &=& c_{ss'}\delta(t_{12}) \Big[   J (2G_{x,ss'}^{\alpha\beta}) +  U (2G_{x,ss'}^{\alpha\beta})^{q/2-1} [M_{(x)}]^{\alpha\gamma}_s [M_{(x)}]^{\beta\delta}_{s'} (2G_{\bar x,ss'}^{\gamma\delta})^{q/2} \Big],
\eea
where we have defined $[M_{(1)}]_s^{\alpha\beta}=M_s^{\alpha\beta}$, $[M_{(2)}]_s^{\alpha\beta}= M_s^{\beta\alpha}$, and $\bar 1 =2 $, $\bar 2 = 1$. Because we are interested in the steady state in the long-time limit, we assume the solution depends only on the time difference. Then the first equation can be solved by Fourier transform,
\bea
    \hat G^{-1}_x(\omega) &=& -i \omega \hat S  - \hat \Sigma_x(\omega),
\eea
where $\hat G_x(t_1, t_2) = \int \frac{d\omega}{2\pi} \hat G_x(\omega) e^{i \omega t_{12} }$, $\hat \Sigma_x(t_1, t_2) = \int \frac{d\omega}{2\pi} \hat \Sigma_x(\omega) e^{i \omega t_{12} }$, with $t_{12}= t_1 - t_2$. The Fourier transform is over the continuous frequency as we take $T\rightarrow \infty$. Moreover, because the self energy is proportional to a Dirac $\delta$ function, $\delta(t_{12})$, in frequency space it is just a constant, $\hat \Sigma_x(\omega) = \hat \Sigma_x$. In the following, we will always use $\hat \Sigma_x $ without argument to denote the constant self energy in frequency space. 

An inspection of the self energy equation of motion,
\bea \label{eq:self-energy}
    \Sigma_{1,-+}^{\alpha\beta} &=&  J (2G_{1,-+}^{\alpha\beta}) + U (2G_{1,-+}^{\alpha\beta})^{q/2-1} \sum_\gamma \epsilon^{\alpha \gamma}  (2G_{2,-+}^{\gamma\beta})^{q/2}, \\
	\Sigma_{2,-+}^{\alpha\beta} &=& J (2G_{2,-+}^{\alpha\beta}) + U (2G_{2,-+}^{\alpha\beta})^{q/2-1} \sum_\gamma (\epsilon^T)^{\alpha \gamma}  (2G_{2,-+}^{\gamma\beta})^{q/2},
\eea
suggests that the solution is given by permutation matrices. Let $\tau$ denote a general $n$-by-$n$ permutation matrix. It is easy to check for any pairs of permutation matrices, $(\tau_A, \tau_{\bar A})$, satisfying $\tau_{A} = \epsilon \tau_{\bar A}$, there is a solution given by the Green's function, $\hat G_A(t_1, t_2) =\hat G(t_1,t_2, \tau_A), \hat \Sigma_A =\hat \Sigma(\tau_A)$, and $\hat G_{\bar A}(t_1, t_2) =\hat G(t_1,t_2, \tau_{\bar A}), \hat\Sigma_{\bar A} = \hat\Sigma(\tau_{\bar A}) $, where the functions are defined by
\bea \label{eq:solution}
    \hat G(t_1, t_2,\tau) = \frac{e^{- \Lambda |t_{12}|}}2  \left( \ba{cccc} \sgn(t_{12}) & -  \tau^T \\  \tau & -\sgn(t_{12})   \ea \right), \quad 
	\hat \Sigma(\tau) =  \Lambda \left( \ba{cccc} 0 & -  \tau^T \\  \tau & 0   \ea \right),
\eea
where $t_{12} = t_1 - t_2$, $\Lambda = J + U$, the solution is written in the Keldysh space and $\tau$ is a $n$-by-$n$ permutation matrix in the replica space. 

To show these are the center-of-mass time invariant solutions, we take the transpose of the first equation and use the antisymmetric property of the bilocal field $\hat G^T = - \hat G$ to get~\cite{stanford2021subleading}
\bea
    \partial_{t_1} \hat S \hat G_x(t_1,t_2) + \partial_{t_2} \hat G_x(t_1,t_2)\hat S  = [\hat \Sigma_x(t_1,t_2), \hat G_x(t_1,t_2)].
\eea
For the Brownian model, we have further $ \hat \Sigma_x(t_1,t_2) = \hat \Sigma_x(t_1) \delta(t_1-t_2)$. Setting $t_1 = t_2$, the differential equation for center-of-mass time is
\bea \label{eq:eom}
    \partial_t \hat S \hat G_x(t,t) \hat S &=& [\hat \Sigma_x(t), \hat G_x(t,t)].
\eea
The right-hand side vanishes for the saddle-point solutions~(\ref{eq:solution}). 

The solution given by~(\ref{eq:solution}) is onshell in the bulk $0<t<T$, but does not satisfy the boundary condition. The full dynamics set by the equation of motion~(\ref{eq:eom}) with the boundary condition~(\ref{eq:bc}) is a complicated problem for general $n$, involving multiple-variable non-linear differential equations for which a general solution does not exist. The solution given by a permutation matrix~(\ref{eq:solution}) is actually a fixed point of the equation of motion in the long-time limit. We show in Appendix~\ref{append:saddle} that the real onshell solution in the long-time limit is given by~(\ref{eq:solution}) in the bulk, with deviations from it near the boundary to account for the correct boundary conditions. The deviation is exponentially suppressed as time moves away from the boundary, with a time scale set by $\frac1{\sqrt{U(2J+U)}}$. In the following, we will see that the solution~(\ref{eq:solution}) gives an essential contribution to the entanglement of the steady state. In Appendix~\ref{append:saddle}, we also argue that the solution~(\ref{eq:solution}) correctly captures the time-independent part of the onshell action at the late time. There are two further remarks on the solution given by~(\ref{eq:solution}) being the relevant one. First, summing over these solutions correctly leads to the Page value in the long-time limit. If there were a contribution due to the deviation near the boundary, the R\'enyi entropy would have involved microscopic parameters like $U/J$, which is not reasonable for a scrambling system at late time. Second, it is reasonable to expect that when the $n \rightarrow 1$ limit is taken, the effect of the boundary conditions on the solutions vanishes as there is nothing to permute at $n = 1$. This is similar to the vanishing of backreaction in the gravity setup~\cite{almheiri2019replica, penington2019replica}. 

Actually, for any choice of $s^{\alpha\beta}_x = \pm$, $\alpha \ne \beta$ (we define $s^{\alpha\alpha}_x =1$ for completeness), there is a center-of-mass time invariant solution given by the modified permutation matrix $\tau_x^{\alpha\beta} \equiv s^{\alpha\beta}_x \tau^{\alpha\beta}$, namely, $G_{x \in A} = G(t_1,t_2, \tau_{A,x})$, $G_{x \in \bar A} = G(t_1,t_2, \tau_{\bar A,x})$, where we have referred to $x=1$ as $x \in A$, and $x=2$ as $x \in \bar A$. The emergence of these solutions is due to the Fermi parity symmetry. The theory~(\ref{eq:LRhamiltonian}) has Fermi parity conservation at each site, whose transformation law is $\psi_{x,j} \rightarrow - \psi_{x,j}, j=1,...,N$. When we extend the theory to $n$ replicas, the symmetry is also extended to independent transformation in each replica, namely, $\psi_{x,\pm}^\alpha \rightarrow - \psi_{x,\pm}^\alpha $ in terms of the Keldysh field~\footnote{Note that the large-$N$ action has an emergent $SO(N)$ symmetry, and we can talk about $\psi_{x,j,s}^\psi$ without referring to an individual $j$.}. The solutions given by $\tau_x^{\alpha\beta}$ can be divided into classes that within each class they are related to each other via these transformations. There are inequivalent classes of solutions distinguished by, 
\bea \label{eq:parity}
    P_x = \prod_{\alpha=1}^n (2i \chi_{x}^\alpha \psi_{x}^\alpha).
\eea
Because of the $SO(N)$ symmetry, the Fermi parity is conserved for each individual flavor. Thus, for each site there are two distinct classes of solutions for each permutation matrix $\tau_x$ distinguished by $P_x = \pm 1$. 

Put it in another way, the saddle-point solution spontaneously breaks the local Fermi parity symmetry, and the parity transformation can bring one solution to the other. Those solutions that can be connected by the transformation is distinguished by $P_x = \pm 1$ in~(\ref{eq:parity}), i.e., there are two inequivalent classes for each site. 
Since the initial state is a product EPR state between each site where for each site, the Fermi parity is $P_x = 1, \forall x$, there is a unique class of solutions that is allowed by the initial state.

\subsection{Summing over all saddle points}

Let us now compute the onshell action~(\ref{eq:action}) by plugging in the saddle-point solution. First notice that the forward and backward evolution cancels because of $c_{ss'}$,
\bea
 \int dt  \sum_{x} \frac{J}{8}  c_{ss'} (2G_{x,ss'}^{\alpha\beta}(t,t))^2 + \frac{U}{2q}   c_{ss'}(2G_{1,ss'}^{\alpha\beta}(t,t))^{q/2} M^{\alpha\gamma}_s M^{\beta\delta}_{s'} (2G_{2,ss'}^{\gamma\delta}(t,t))^{q/2} = 0.
\eea
There is also an exponential decaying part,
\bea \label{eq:decay}
    \exp\left(- \frac12 \int dt \sum_x  \Sigma_{x,ss'}^{\alpha\beta} G_{x,ss'}^{\alpha\beta}(t,t)\right) = \exp(- n \Lambda T ).
\eea

We discuss the Pfaffian in detail. The evaluation of the Pfaffian can be mapped to a Kitaev chain problem. Because the large-$N$ action~(\ref{eq:action}) has an $SO(N)$ symmetry, the Pfaffian at each site (we first suppress the site index, and later restore it when we are familiar with the evaluation for a single site) is equivalent to a transition amplitude of an EPR state of an effective single flavor Majorana fermion for two contours of each replica, $\psi_\pm^\alpha$. The EPR state is $(\psi^{\alpha} + i \chi^\alpha )|\text{EPR} \rangle = 0$, $\forall \alpha = 1,...,n$, and the Pfaffian is
\bea
    \Pf \Big[  \hat S \partial_t  - \Sigma \Big] = \langle \text{EPR} | e^{ T H}  |\text{EPR} \rangle, \quad H = \frac12 (\psi, \chi) \left( \ba{cccc} 1 & 0 \\ 0 & i \ea \right) \Sigma \left( \ba{cccc} 1 & 0 \\ 0 & i \ea \right) \left( \ba{cccc} \psi \\ \chi \ea \right).
\eea
Here the self-energy $\Sigma$ is given by some $n$-by-$n$ permutation matrix. 

Now, all permutations of $n$ elements can be decomposed into cyclic permutations (or cycles) with length $n_{\check\tau_{(i)}}$ such that $\sum_i n_{\check\tau_{(i)}} = n$. We use a check over the letter, $\check{\tau}$, to denote cycles. Here since the problem is quadratic, we can indeed simplify the Hamiltonian by taking the permutation matrix $\tau$ into block diagonal individual cycles $\tau = \oplus_{i} \check \tau_{(i)}$, and for each of these cycles, we can bring them to the following canonical form,
\bea \label{eq:canonical}
    \check\tau_{(i)}^{\alpha\beta} =  \begin{cases} \delta^{\alpha+1,\beta}, & n_{\check\tau_{(i)}} = \text{odd} \\
                            \sgn(\beta-\alpha)\delta^{\alpha+1,\beta}, & n_{\check\tau_{(i)}} = \text{even} \end{cases},
\eea
where we defined $\delta^{n+1,\beta} = \delta^{1,\beta}$. Note that for the even length cycles, it has an additional minus sign to be consistent with the even Fermi parity $P_x = 1$.

Now the Hamiltonian decomposes into individual cycles. Let us consider a cycle $\check \tau$ in the canonical form with length $n_{\check\tau}$. We define
\bea
    H(\check\tau) = \frac{\Lambda}2  (\psi, \chi) \left( \ba{cccc} 0 & -  i \check\tau^T \\  i \check\tau & 0   \ea \right) \left( \ba{cccc} \psi \\ \chi \ea \right),
\eea
which is nothing but a Kitaev chain~\cite{kitaev2001unpaired} with length $n_{\check\tau}$. Using the complex fermion, 
\bea \label{eq:complex}
    c_\alpha = \frac{\psi^\alpha + i \chi^\alpha}{\sqrt2}, \quad c_\alpha^\dag = \frac{\psi^\alpha - i \chi^\alpha}{\sqrt2},
\eea
satisfying $\{c_\alpha, c_\beta^\dag\} = \delta_{\alpha\beta}$, we can bring the Hamiltonian into
\bea
    H(\check\tau) &=& \frac{\Lambda}2 \sum_{\alpha=1}^{n_{\check \tau}} [( c_\alpha^\dag c_{\alpha+1} +  c_\alpha c_{\alpha +1 } + h.c.) + ((-1)^{n_{\check \tau}+1} c_n c_1^\dag + h.c.)] \\
    &=&  \Lambda \sum_{k} (c_k^\dag , c_{-k}) \left( \ba{cccc} - \cos k & -i \sin k \\
    i \sin k & \cos k \ea \right) \left( \ba{cccc} c_k \\ c_{-k}^\dag \ea \right)  + \text{mod}(n_{\check \tau},2) \Lambda (c_0^\dag c_0 + \frac12),
\eea
where we have the BdG Hamiltonian~\cite{altland2010condensed} in the second line. The choice of momentum needs some explanation because the odd and even lengths have different boundary conditions. When $n_{\check \tau}$ is an odd integer, the fermion satisfies periodic boundary condition, $k = \frac{2j\pi}{n_{\check \tau}}$, and the summation is over $j=1,...,\frac{n_{\check \tau}-1}2$. When $n_{\check \tau}$ is an even integer, the fermion satisfies anti-periodic boundary condition, $k = \frac{(2j-1)\pi }{n_{\check \tau}}$, and the summation is over $j=1,...,\frac{n_{\check \tau}}2$. The last term with zero momentum in the second line appears only for odd integer $n_{\check \tau}$ because it is allowed under periodic boundary conditions. 

It is easy to diagonalize the Hamiltonian with the Bogoliubov quasiparticle operators,
\bea
    d_{-,k} &=&  - i \cos \frac{k}2 c_k + \sin \frac{k}2 c_{-k}^\dag \\
    d_{+,k} &=&  i \sin \frac{k}2 c_k + \cos \frac{k}2  c_{-k}^\dag, 
\eea
such that
\bea
    H(\check \tau) &=& \Lambda \sum_{k} (d_{+,k}^\dag d_{+,k} - d_{-,k}^\dag d_{-,k}) + \text{mod}(n_{\check \tau},2) \Lambda (c_0^\dag c_0 + \frac12).
\eea
Each of the eigenmodes is independent for the choice of momentum discuss in above, and the exponential of the Hamiltonian is given by
\bea
    e^{TH(\check \tau)} &=& e^{\text{mod}(n_{\check \tau},2) \frac{\Lambda}2 } \left( 1+ (e^{\Lambda T}-1) c_0^\dag c_0 \right) \\ && \times \prod_{k} \left( 1+  (e^{\Lambda T} -1) d_{+,k}^\dag d_{+,k} + 1+  (e^{-\Lambda T} -1) d_{-,k}^\dag d_{-,k} \right) \\
    & \ni&  e^{\text{mod}(n_{\check \tau},2) \frac{\Lambda}2 } \prod_{k} e^{\Lambda T} \cos^2 \frac{k}2 \, c_{k} c_{k}^\dag,
\eea
where because the EPR state is the vacuum of a complex fermion~(\ref{eq:complex}), we express the Bogoliubov quasiparticle using the complex fermion and show the largest nonvanishing component for each $k$ in the second line. Then it is straightforward to evaluate the transition amplitude of the EPR state,
\bea
    \langle \text{EPR} | e^{TH(\check\tau)} |\text{EPR} \rangle = e^{ \frac{n_{\check \tau}}2 \Lambda T} \prod_{k} \cos^2 \frac{k}2 =  e^{ \frac{n_{\check \tau}}2 \Lambda T} 2^{-n_{\check \tau} + 1}. 
\eea
where the range of product of $k$ is the same as the summation we discussed before.

We now know the contribution for a cycle $\check\tau$. It is time to discuss the full solution including different sites. For the full solution, recall that the solution in $x \in A$ and $x \in \bar A$ are related by the cyclic permutation $\epsilon$, namely, $\tau_{A} = \epsilon \tau_{\bar A}$. Suppose that they can be decomposed into $m_\text{cyc}$ number cycles, i.e., those two permutation matrices can be built using $\check \tau_{(i)}$, $i=1,2,..., m_\text{cyc}$. The sum of the length of each cycles should satisfy $\sum_{j=1}^{m_\text{cyc}} n_{\check\tau_{(j)}} = 2n$. Then the Pfaffian becomes
\bea \label{eq:pfaffian}
    \prod_{x=1,2} \Pf \Big[  \hat S \partial_t  - \Sigma_x \Big] =  \langle \text{EPR} | e^{T \sum_{i=1}^{m_\text{cyc}} H(\check\tau_{(i)})} |\text{EPR} \rangle = \prod_{i=1}^{m_\text{cyc}} e^{ \frac{n_{\check \tau}}2 \Lambda T} 2^{-n_{\check \tau} + 1} =   e^{ n \Lambda T} 2^{-2n + m_\text{cyc}} . \nn \\
\eea
Because $n$ is the number of replicas, it is natural to expect the action to have a factor of $n$. What is interesting is that the Pfaffian depends on the number of cycles in the solution, $m_\text{cyc}$, and does not depend on other details of the permutation matrix.

For a pair of permutations, $(\sigma,\tau)$, that are related by $\sigma = \epsilon \tau$, it is known that the maximal possible number of cycles that can be decomposed from them is $m_\text{cyc} = n+1$~\cite{jacques1968genre,dulucq1998combinatorial}.  
Then we have
\bea 
    \prod_{x=1,2} \Pf \Big[  \hat S \partial_t  - \Sigma_x \Big] = \langle \text{EPR} | e^{T \sum_{i=1}^{n+1} H(\check\tau_{(i)})} |\text{EPR} \rangle =   e^{ n \Lambda T} 2^{-n + 1} .
\eea
The exponential growing in time part cancels the decaying part in~(\ref{eq:decay}), and the onshell action becomes $e^{-I_\text{onshell}} = e^{(1-n)N \log 2}$.

For those pairs with less cycles, the contribution will be exponentially suppressed, i.e., $2^{N m_\text{cyc}}$, so we ignore them. There are $C_n$ pairs of permutation matrices, $(\sigma,\tau)$, that are related by $\sigma = \epsilon \tau$, having the maximal number of cycles, where $C_n$ is the Catalan number~\cite{jacques1968genre,dulucq1998combinatorial}.  In addition to that, we have also a degeneracy coming from the local Fermi parity symmetry $\psi^\alpha_{j} \rightarrow - \psi^\alpha_{j}, j=1,...,N$ independently for each replica $\alpha = 1,..,n$, which leads to a number $(2^n/2)^2 = 2^{2(n-1)}$~\footnote{The $1/2$ inside the parentheses is due to the fact that the Fermi parity of the initial state is even at each site, so that only half of solutions are allowed, while the $2$ in the power is because we have two sites.}. Taking into account these degenerate saddle points, the steady state R\'enyi entropy is given by
\bea \label{eq:renyi}
    S_A^{(n)} = (N-2) \log 2 + \frac1{1-n} \log C_n.
\eea
To get the von Neumann entanglement entropy, it is useful to represent the Catalan number as $C_n = \frac1{2\pi} \int_0^4 dx x^{n-1} \sqrt{(4-x)x}$. We can analytically continue the number to $n \rightarrow 1$, and this leads to
\bea
    \lim_{n\rightarrow 1} \frac1{1-n} \log C_n &=& \lim_{n\rightarrow 1} \frac1{1-n} \log \left(\frac1{2\pi} \int_0^4 dx  \sqrt{(4-x)x} -(1-n) \frac1{2\pi} \int_0^4 dx \log x \sqrt{(4-x)x} \right) \nn \\
    &=& \lim_{n\rightarrow 1} \frac1{1-n} \log  (1 - \frac12(1-n) ) = - \frac12.
\eea
so the von Neumann entanglement entropy is $S_A = N \log 2 - 2\log 2 - \frac12$. The first $1/N$ is due to the Fermi parity symmetry, i.e., starting from an EPR state the steady state can explore half of the total Hilbert space at each site~\cite{jian2020note}. The second $1/N$ correction is consistent with the Page value in the limit of large Hilbert space dimension~\cite{page1993average}.

We can do better than just calculating the von Neumann entanglement entropy. This explicit form~(\ref{eq:renyi}) allows us to calculate the entanglement spectrum density of states. This is done by considering the resolvant of reduced density matrix of subsystem $A$~\cite{penington2019replica},
\bea
    R(\lambda) = \frac{1}{\lambda - \rho_A} = \frac1{\lambda} + \sum_{k=1}^\infty \rho_A^k, \quad \rho_A = \Tr_{\bar A} \rho.
\eea
The entanglement spectrum density of states is given by the resolvant through~\footnote{As we know the steady state can only explore the Hilbert space with even Fermi parity at each site, we restrict the trace to that subspace, e.g., $\Tr[1] = 2^{N-2}$. }
\bea
    D(\lambda) = \frac1{2\pi i} \lim_{\delta \rightarrow 0} (\Tr[R(\lambda - i \delta)] - \Tr[R(\lambda + i \delta)]).
\eea
The trace of resolvant can be obtained by the saddle-point action,
\bea
    \Tr[R(\lambda)] &=& \frac1{\lambda} 2^{N-2}  + 
    \sum_{k=1}^\infty \frac{\Tr[\rho_A^k(\lambda)]}{\lambda^{k+1}} 
    = \frac1{\lambda} 2^{N-2}  + 
    \sum_{k=1}^\infty \frac{\Tr[\rho^{\otimes k} M_\text{cyc}(A)]}{\lambda^{k+1}} \\
    &=& \frac1{\lambda} 2^{N-2}  + 
    \sum_{k=1}^\infty \frac{C_k e^{(1-k)(N-2) \log 2}}{\lambda^{k+1}} 
    = 2^{2N-5} \left( 1- \sqrt{1- \frac{2^{4-N}}\lambda} \right),
\eea
where in the third equality we used the saddle-point results and in the last equality we performed the summation. The density of state function becomes
\bea
    D(\lambda) = \frac{2^{2N-5}}{\pi} \frac{\sqrt{\lambda(2^{4-N}-\lambda)}}{\lambda},
\eea
which is equivalent to the result of a random pure state~\cite{page1993average} in the subsector accessible from the initial EPR state. It is expected since unitary evolution generated from the Brwonian SYK Hamiltonian approaches the Haar random unitary at late time~\cite{jian2021quantum, stanford2021subleading}. This result strongly supports that the saddle-point solutions given by the permutation matrix~(\ref{eq:solution}) play an essential role in late-time entanglement entropy.

If there is no cyclic permutation operator in the path integral, we are just calculating the $n$-th power of the trace of a density matrix. In this case, the solutions described by the permutation matrix also exists. However, the relation between solution in $A$ and $\bar A$ changes to $\tau_A = \tau_{\bar A}$, so the maximal decomposition to cycles is given by the identity matrix, where each diagonal element is a trivial cycle with length one. Taking $\tau_A$ and $\tau_{\bar A}$ to be the identity matrix where the number of cycles is $m_\text{cyc}=2n$ in total, we have from~(\ref{eq:pfaffian})
\bea \label{eq:denominator}
     \prod_{x=1,2} \Pf \Big[  \hat S \partial_t  - \Sigma_x \Big] =  e^{ n \Lambda T}. 
\eea
The exponential growing part gets cancelled exactly by~(\ref{eq:decay}), so we get one, which is $n$-th power of the trace of a density matrix.

Finally, although here we only consider two coupled clusters, it is straightforward to generalize the calculation to the half-chain quasi entropy of a chain with $L$ sites. The calculation is the same: the solution in $A$ and $\bar A$ are given by pairs of permutation matrices, $(\tau_A,\tau_{\bar A})$ with $\tau_A = \epsilon \tau_{\bar A}$, where now $A$ is half of the chain and $\bar A$ is the rest. 
Taking into account the length of the chain and of the subregion $A$, the half-chain entanglement is given by $S_A = \frac{NL}2 \log 2 - L \log 2 - \frac12 $, where the first $1/N$ is due to the Fermi parity symmetry at each site, i.e., starting from the EPR state, the steady state can only explore the subspace of even Fermi parity at each site.

\section{Entanglement phase transition in the monitored Brownian SYK chains} \label{sec:monitor}

In this section, we will consider the monitored Brownian SYK chains introduced in Section~\ref{sec:model}. We focus on the effect of the measurement on the saddle-point solutions and thus the quasi entropy. Then we take analytical continuation $n \rightarrow 1$ to get the von Neumann entanglement entropy for the cyclic symmetric solution.

\subsection{Saddle-point analysis}
To generalize the story in the coupled Brownian SYK clusters into monitored Brownian SYK chains, we double the degrees of freedom into left and right chains, and add the effective action description~(\ref{eq:monitor}) for continuous monitoring. Similar to the unitary case, we will consider EPR initial states. It is described by two copies of the fermions, denoted by $\psi_{x,a,j}$ and $\chi_{x,a,j}$ at every site in both chains. The initial density matrix is given by
\bea
    \rho = |\text{EPR} \rangle  \langle \text{EPR} |, \quad    (\psi_{x,a,j} + i \chi_{x,a,j} )| \text{EPR} \rangle =0, \\
    \quad \forall x=1,...,L, \quad  \forall a = L ,R, \quad  \forall j=1,...,N. 
\eea
We divide the chain into $A$ and $\bar A$, and compute the quasi entropy of the steady state~(\ref{eq:quasi_disorder}). The only modification of the quasi entropy is the presence of monitoring, and we have shown that the effect of monitoring part is described by~(\ref{eq:monitor}). Note that the redefinition~(\ref{eq:redefine}) will not change the form of the effective action~(\ref{eq:monitor}). It is then straightforward to get the $G$-$\Sigma$ action for $\mathbb E \Tr[\tilde \rho^{\otimes n} M_\text{cyc}(A)] = \int DG D\Sigma e^{-I(A)} $,
\bea \label{eq:action2}
	&& - \frac{I(A)}{N} = \log \Pf [ \hat S \partial_t  - \hat \Sigma_{x}] + \int dt_1 dt_2  \Big( - \frac{1}2  \Sigma_{x,ab,ss'}^{\alpha\beta} G_{x,ab,ss'}^{\alpha\beta} + \frac{J}{8} \delta(t_{12}) \delta_{ab} c_{ss'} (2G_{x,ab,ss'}^{\alpha\beta}(t_1,t_2))^2 \Big)  \nn\\
	&& + \frac{U}{4q} \int dt_1 dt_2 \delta(t_{12}) \delta_{ab} c_{ss'}(2G_{x,ab,ss'}^{\alpha\beta}(t_1,t_2))^{q/2} M^{\alpha\gamma}_s(x,x+1) M^{\beta\delta}_{s'}(x,x+1) (2G_{x+1,ab,ss'}^{\gamma\delta}(t_1,t_2))^{q/2}  \nn \\
	&& + \frac{i \mu}2 \int dt_1 dt_2 \delta(t_{12}) \delta_{ss'} \delta^{\alpha\beta} (\delta_{aL}\delta_{bR} -\delta_{aR}\delta_{bL}) G_{x,ab,ss'}^{\alpha\beta}(t_1,t_2),
\eea
where the summation over the site indices, the chain indices, the replica indices, and the contour indices, $x, a, b, \alpha, \beta, \gamma, \delta, s,s'$, is implicit. The third line is the monitoring part from~(\ref{eq:monitor}). In the second line, $M_s(x,y)$ is the cyclic permutation operator at the boundary of intervals $A$ and $\bar A$,
\bea
    M^{\alpha\beta}_+(x,y) &=& \delta^{\alpha\beta}, \\
    M^{\alpha\beta}_-(x,y) &=& \begin{cases}
        \delta^{\alpha\beta} & x, y \in A \text{ or } x, y \in \bar A \\
        \epsilon^{\alpha\beta} & x \in A \text{ and } y \in \bar A \\
        \epsilon^{\beta\alpha} &  x \in \bar A \text{ and } y \in A
    \end{cases},
\eea
which is only nontrivial when $x$ and $y$ are located at different subsystems, and is introduced to calculate the quasi entropy between $A$ and $\bar A$. We assume the length of $A$ and $\bar A$ are equal and given by $L/2$. The boundary conditions of the fields are similar to~(\ref{eq:bc}) with a trivial extension to left and right chains.

The equations of motion from the action read
\bea
     \hat G^{-1}_x &=& \hat S \partial_t - \hat \Sigma_x, \\
     \Sigma_{x,ab,ss'}^{\alpha\beta} &=&  \delta(t_{12})  \Big[   J \delta_{ab} c_{ss'} (2G_{x,ab,ss'}^{\alpha\beta})  + i \mu \delta_{ss'} \delta^{\alpha\beta} (\delta_{aL}\delta_{bR} -\delta_{aR}\delta_{bL}) \nn \\
    && +  \frac{U}2 \delta_{ab} c_{ss'} \Big((2G_{x,ab,ss'}^{\alpha\beta})^{q/2-1} M^{\alpha\gamma}_s(x,x+1) M^{\beta\delta}_{s'}(x,x+1) (2G_{x+1, ab ,ss'}^{\gamma\delta})^{q/2} \nn \\
    && + (2G_{x-1,ab,ss'}^{\gamma\delta})^{q/2} M^{\gamma\alpha}_s(x-1,x) M^{\delta\beta}_{s'}(x-1,x) (2G_{x,ab,ss'}^{\alpha\beta})^{q/2-1} \Big) \Big]. 
\eea
We assume that the solutions are uniform in $A$ and $\bar A$ up to the parity transformation. The invariant quantity that distinguishes different classes of solutions now becomes the total Fermi parity of left and right Majorana fermions at each site, $P_{x,L} P_{x,R}$. In the presence of measurement, only the total parity of left and right Majorana fermions at each site is conserved, i.e., the theory~(\ref{eq:action2}) is invariant under transformation that acts on both left and right Majorana fermoins, $\psi_{x,L/R,\pm}^{\alpha} \rightarrow -\psi_{x,L/R,\pm}^{\alpha}$. 

The solution is now modified by the presence of the monitoring. Similar to the discussion in the previous section, it is easy to check that the ansatz $G_{x \in A}= G(t_1,t_2,\tau_{A,x})$, $\Sigma_{x\in A}(t_1,t_2) = \Sigma(\tau_{A,x})$, and $G_{x \in \bar A}(t_1,t_2)= G(t_1,t_2,\tau_{\bar A,x})$, $\Sigma_{x \in \bar A} = \Sigma(\tau_{\bar A,x})$ solves the equation of motion, with $\tau_A = \epsilon \tau_{\bar A}$,  where $G(t_1,t_2,\tau)$ and $\Sigma(\tau)$ are now given by
\bea
\label{eq:solution2}   && G(t_1,t_2,\tau) = \frac{e^{-\Lambda |t_{12}|}}{2}  \left( \ba{cccc} \sgn(t_{12}) & -i\sin\theta & -\cos\theta\, \tau^T & 0 \\
                            i\sin\theta & \sgn(t_{12}) & 0 & - \cos\theta\, \tau^T \\
                            \cos\theta\, \tau & 0 & -\sgn(t_{12}) & -i \sin\theta \\
                            0 & \cos\theta\, \tau & i\sin\theta & -\sgn(t_{12}) \ea \right), \\
   && \Sigma(\tau) = \left( \ba{cccc} 0 & i\mu & -\Lambda \tau^T & 0 \\
                            -i\mu & 0 & 0 & - \Lambda \tau^T \\
                            \Lambda \tau & 0 & 0 & i \mu\\
                            0 & \Lambda \tau & -i\mu & 0 \ea \right),
\eea
The solution is written in the $\left(\psi_{L,+}^\alpha,\psi_{R,+}^\alpha,\psi_{L,-}^\alpha,\psi_{R,-}^\alpha \right)$ basis, and for the Green's function, we have introduced $\tan \theta = \frac{\mu}{\Lambda}$, such that when $\theta=0$, it reduces to the unmeasured case. Recall that $\tau_x^{\alpha\beta} = s_x^{\alpha\beta} \tau^{\alpha\beta} $ is the permutation matrix that characterizes the correlation between different replicas. The EPR initial state allows the solutions with $P_{x,L} P_{x,R} = 1$, but only the permutation solution for $P_{x,L} = P_{x,R} = 1$ can smoothly connect to the zero measurement limit. Note that in the non-measurement case, the Fermi parity at two chains is conserved separately. So, we consider this sector of solutions, which includes $(2^n/2)^L = 2^{L(n-1)}$~\footnote{Because the parity transformation acts simultaneously on left and right Majorana fermions, we have a similar degeneracy as before.} degenerate solutions from the parity transformation. 

\begin{figure}
    \centering
    \includegraphics[width=0.4\textwidth]{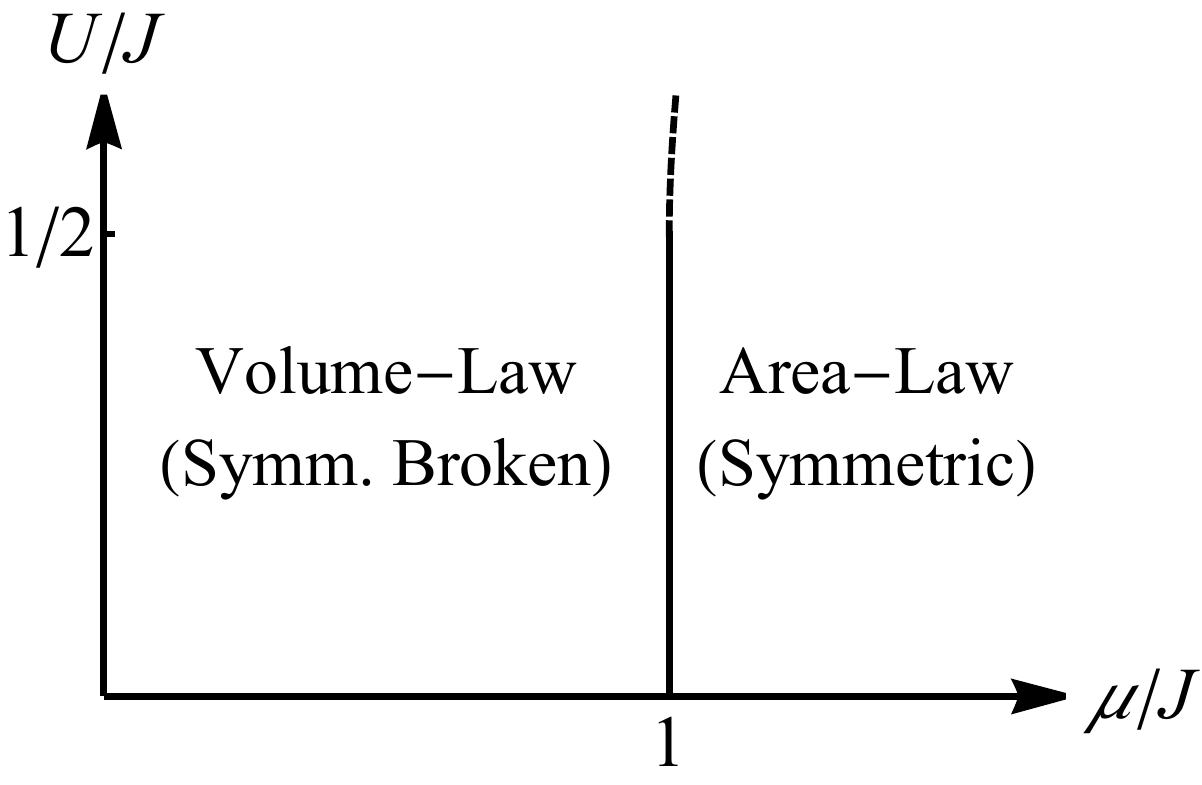}
    \caption{The phase diagram as a function of measurement strength for $q=4$. The phase boundary is determined by (\ref{eq:phase}). The solid line indicates a continuous transition, while the dashed line indicates a first-order transition.}
    \label{fig:phase}
\end{figure}

In the presence of measurement, the parameter $\Lambda$ satisfies 
\bea \label{eq:phase}
    \Lambda = U \left( \frac{\Lambda}{\sqrt{\Lambda^2 + \mu^2}} \right)^{q-1} + J \frac{\Lambda}{\sqrt{\Lambda^2 + \mu^2}}.
\eea
For small interaction strength $U \ll J$, $\Lambda$ is given by
\bea \label{eq:lambda}
    \Lambda = J \left(1- \frac{\mu^2}{J^2} \right)^{\frac12} +  U \left( 1 - \frac{\mu^2}{J^2} \right)^{\frac{q-3}2} + \mathcal{O}\left( U^2 \right).
\eea
$\Lambda$ vanishes at $\mu = J$. At this point $\theta = \frac{\pi}2$, and the correlation between left and right chains is maximal while the correlation between different replicas vanishes. The solution restores the replica symmetry. We will see in the following section that this corresponds to an entanglement transition, namely, when the replica symmetry is broken, the von Neumann entanglement entropy satisfies a volume law, while when the replica symmetry is unbroken, the von Neumann entanglement entropy satisfies an area law. A plot of the phase diagram for $q=4$ as a function of the monitoring rate is shown in Fig.~\ref{fig:phase}.

\subsection{Quasi entropy by summing over all saddle points}

Now we evaluate the onshell action on the above solution. Due to the presence of monitoring, the forward and backward evolution no longer cancel,
\bea \label{eq:onshell_unitary}
    && \int dt \Big[\frac{U}{4q}  \delta_{ab} c_{ss'}(2G_{x,ab,ss'}^{\alpha\beta}(t,t))^{q/2} M^{\alpha\gamma}_s(x,x+1) M^{\beta\delta}_{s'}(x,x+1) (2G_{x+1,ab,ss'}^{\gamma\delta}(t,t))^{q/2} \nn \\
    && + \frac{J}{8} \delta_{ab} c_{ss'} (2G_{x,ab,ss'}^{\alpha\beta}(t,t))^q\Big]  = \left[\frac{J}2 \Big[ \Big(\frac{\Lambda}{\sqrt{\Lambda^2 + \mu^2}} \Big)^2 -1 \Big]+ \frac{U}{q} \Big[ \Big(\frac{\Lambda}{\sqrt{\Lambda^2 + \mu^2}} \Big)^q -1\Big] \right] n L T.  \nn\\
\eea
From this expression, we can check that when the monitoring strength is zero, $\mu=0$, the contributions from the forward and backward evolution do cancel. The monitoring part itself also leads to a contribution,
\bea \label{eq:onshell_monitor}
    && \int dt \frac{i \mu}2 \delta_{ss'} \delta^{\alpha\beta} (\delta_{aL}\delta_{bR} -\delta_{aR}\delta_{bL}) G_{x,ab,ss'}^{\alpha\beta}(t,t) = - \frac{\mu^2}{\sqrt{\Lambda^2 + \mu^2}} n L T.
\eea
Similar to the unitary case, we have also the following contribution,
\bea \label{eq:decay2}
    - \frac12 \int dt   \Sigma_{x,ab,ss'}^{\alpha\beta} G_{x,ab,ss'}^{\alpha\beta}(t,t) = - \sqrt{\Lambda^2 + \mu^2 } n L T.
\eea
In the unitary case, this exponential decaying part is cancelled by the Pfaffian. We will see that it is also true for monitoring case. Nevertheless, the other parts~(\ref{eq:onshell_unitary}) and~(\ref{eq:onshell_monitor}) cannot be canceled. This is consistent with $\tilde \rho_{\nu}$ being an unnormalized density matrix.


The evaluation of the Pfaffian can again be mapped to a transition amplitude of Kitaev chains, but now we have two coupled Kitaev chains. Similarly, because the large-$N$ action~(\ref{eq:action2}) has an $SO(N)$ symmetry, the Pfaffian is equivalent to the transition amplitude of an EPR state of an effective single flavor Majorana fermion for two contours of each replica, $\psi_{x,a,s}^\alpha$. The EPR state is $(\psi_{x,a}^{\alpha} + i \chi_{x,a}^\alpha )|\text{EPR} \rangle = 0$, $\forall x=1,...,L$, $\forall \alpha = 1,...,n$, and $\forall a=L,R$. The Pfaffian is given by the following transition amplitude,
\bea
    \prod_{x=1}^L \Pf \Big[  \hat S \partial_t  - \Sigma_x \Big] &=& \langle \text{EPR} | e^{ T \sum_{x=1}^L H_x}  |\text{EPR} \rangle,
\eea
with the Hamiltonian given by
\bea
    H_x &=& \sum_{a=L,R} \frac12 (\psi_{x,a},  \chi_{x,a}) \left( \ba{cccc} 1 & 0 \\ 0 & i \ea \right) \Sigma_x \left( \ba{cccc} 1 & 0 \\ 0 & i \ea \right) \left( \ba{cccc} \psi_{x,a}  \\ \chi_{x,a} \ea \right).
\eea
Similar to the unitary case, the self-energy $\Sigma_x$ is given by some permutation matrix which can be decomposed into cycles with length $n_{\check\tau_{(i)}}$ such that $\sum_i n_{\check\tau_{(i)}} = n$, and for each of cycles, we can bring them to the canonical form~(\ref{eq:canonical}).

Consider a cyclic permutation $\check\tau$ at one site (so we suppress the site index), by extending the previous calculation to left and right chains, the Pfaffian can be mapped to the amplitude of EPR state with the following Hamiltonian
\bea \label{eq:cycle}
    H(\check\tau) = \frac12 (\psi_{L}, \chi_{L}, \psi_{R}, \chi_{R}) \left( \ba{cccc} 0 & -i \Lambda \check\tau^T & i\mu & 0 \\
                                     i\Lambda \check\tau  & 0 & 0 & -i \mu \\
                                     -i \mu & 0 & 0 & -i \Lambda \check\tau^T \\
                                     0 & i \mu & i\Lambda \check\tau & 0  \ea \right) \left( \ba{cccc} \psi_{L} \\ \chi_{L} \\ \psi_{R} \\ \chi_{R} \ea \right).
\eea
Using the complex fermion,
\bea \label{eq:complex2}
 c_{a,\alpha} = \frac{\psi^\alpha_{a} + i \chi^\alpha_{a}}{\sqrt2}, \quad c_{a,\alpha}^\dag = \frac{\psi^\alpha_{a} - i \chi^\alpha_{a}}{\sqrt2},
\eea
satisfying $\{c_{a,\alpha}, c_{b,\beta}\} = \delta_{ab} \delta_{\alpha\beta}$, we can bring the Hamiltonian to
\bea
    H(\check\tau) &=& \frac{\Lambda}2 \sum_{a,\alpha} ( c_{a,\alpha}^\dag c_{a,\alpha+1} +  c_{a,\alpha} c_{a,\alpha +1 } + h.c.) +  \frac\mu2 \sum_\alpha (i c_{L,\alpha} c_{R,\alpha} + i c_{L,\alpha}^\dag c_{R,\alpha}^\dag + h.c. ) \\
    &=&   \sum_{k} (c_{L,k}^\dag , c_{L,-k}, c_{R,k}^\dag, c_{R,-k}) 
    \left( \ba{cccc} - \Lambda \cos k & -i \Lambda \sin k & 0 & i\mu \\
    i \Lambda \sin k & \Lambda \cos k &  i \mu & 0 \\
    0 & -i \mu & - \Lambda \cos k & -i \Lambda \sin k \\
    -i\mu & 0 & i \Lambda \sin k & \Lambda \cos k \ea \right)
    \left( \ba{cccc} c_{L,k} \\ c_{L,-k}^\dag \\ c_{R,k} \\ c_{R,-k}^\dag \ea \right). \nn \\
\eea
Because the odd and even length have different boundary conditions, the choice of momentum is: when $n_{\check \tau}$ is an odd integer, the fermion satisfies periodic boundary condition, $k = \frac{2j\pi}{n_{\check \tau}}$, and the summation is over $j=0,1,...,\frac{n_{\check \tau}-1}2$. When $n_{\check \tau}$ is an even integer, the fermion satisfies anti-periodic boundary condition, $k = \frac{(2j-1)\pi }{n_{\check \tau}}$, and the summation is over $j=1,...,\frac{n_{\check \tau}}2$. 

The Bogoliubov quasiparticle annihilation operators~\cite{altland2010condensed} are given by
\bea
    d_{L,-,k} &=&  \cos\frac\theta2 \left( i\cos\frac{k}2 c_{L,k} - \sin\frac{k}2 c_{L,-k}^\dag \right) + \sin\frac\theta2 \left( i \sin\frac{k}2 c_{R,k}+ \cos\frac{k}2 c_{R,-k}^\dag \right), \\
    d_{R,-,k} &=&  \sin\frac\theta2 \left( i\sin\frac{k}2 c_{L,k} + \cos\frac{k}2 c_{L,-k}^\dag \right) + \cos\frac\theta2 \left( -i \cos\frac{k}2 c_{R,k}+ \sin\frac{k}2 c_{R,-k}^\dag \right), \\
    d_{L,+,k} &=&  \cos\frac\theta2 \left( -i\sin\frac{k}2 c_{L,k} - \cos\frac{k}2 c_{L,-k}^\dag \right) + \sin\frac\theta2 \left( -i \cos\frac{k}2 c_{R,k}+ \sin\frac{k}2 c_{R,-k}^\dag \right), \\
    d_{R,+,k} &=&  \sin\frac\theta2 \left( -i\cos\frac{k}2 c_{L,k} + \sin\frac{k}2 c_{L,-k}^\dag \right) + \cos\frac\theta2 \left( i \sin\frac{k}2 c_{R,k}+ \cos\frac{k}2 c_{R,-k}^\dag \right),
\eea
where $\tan \theta = \frac{\mu}\Lambda $, and in terms of these operators, the Hamiltonian becomes diagonal,
\bea
    H(\check\tau) &=& \sqrt{\Lambda^2 + \mu^2} \sum_{a=L,R;k>0} (d_{a,+,k}^\dag d_{a,+,k} - d_{a,-,k}^\dag d_{a,-,k}) \\
    && + \text{mod}(n_{\check \tau},2) \sqrt{\Lambda^2 + \mu^2} (d_{L,+,0}^\dag d_{L,+,0} - d_{R,-,0}^\dag d_{R,-,0}).
\eea
Note that only the odd $n_{\check \tau}$ can have zero momentum mode because even $n_{\check \tau}$ should satisfies the anti-periodic boundary condition. With the help of the Bogoliubov quasiparticle operator, the exponential of the Hamiltonian can be simplified by
\bea
    e^{TH(\check\tau)} &=& \left( 1+  (e^{\sqrt{\Lambda^2 + \mu^2}T} -1) d_{L,+,0}^\dag d_{L,+,0} + 1+  (e^{-\sqrt{\Lambda^2 + \mu^2}T} -1) d_{R,-,0}^\dag d_{R,-,0} \right)^{\text{mod}(n_{\check \tau},2)} \nn \\
    && \times \prod_{a=L,R;k > 0} \left( 1+  (e^{\sqrt{\Lambda^2 + \mu^2}T} -1) d_{a,+,k}^\dag d_{a,+,k} + 1+  (e^{-\sqrt{\Lambda^2 + \mu^2}T} -1) d_{a,-,k}^\dag d_{a,-,k} \right) \nn \\
    & \ni&  \left(e^{\sqrt{\Lambda^2 + \mu^2}T} \cos^2 \frac\theta2 c_{L,0} c_{L,0}^\dag \right)^{\text{mod}(n_{\check \tau},2)} \nn \\
    && \times \prod_{a=L,R;k>0} e^{\sqrt{\Lambda^2 + \mu^2} T} \left(\cos^2 \frac\theta2 \cos^2\frac{k}2 + \sin^2 \frac\theta2 \sin^2\frac{k}2 \right) c_{a,k} c_{a,k}^\dag .
\eea
where in the first line we explicit show the contribution of $k=0$ sector that is present for odd $n_{\check \tau}$ only. Remember the transition amplitude is between the vacuum state of the original complex fermion~(\ref{eq:complex2}), so we only show the dominant contribution for each $k$ in the second equality. Therefore, the transition amplitude for a cycle $\check\tau$ becomes
\bea
    \langle \text{EPR} | e^{TH(\check\tau)} | \text{EPR} \rangle = e^{ n_{\check \tau} \sqrt{\Lambda^2 + \mu^2}T} \prod_{k} \left(\cos^2 \frac\theta2 \cos^2\frac{k}2 + \sin^2 \frac\theta2 \sin^2\frac{k}2 \right),
\eea
Here the choice of momentum in the product is that for an odd $n_{\check \tau}$, $k = \frac{2j\pi}{n_{\check \tau}}$, with $j=-\frac{n_{\check \tau}-1}2,...,-1,0,1,...,\frac{n_{\check \tau}-1}2$, and for an even $n_{\check \tau}$, $k = \frac{(2j-1)\pi }{n_{\check \tau}}$, with $j=-\frac{n_{\check \tau}}2+1,...,-1,0,1,...,\frac{n_{\check \tau}}2$, by noticing each term in the product is an even function of $k$.
The product is evaluated to be~(see Appendix~\ref{append:product})
\bea 
    &&\prod_{k} \left(\cos^2 \frac\theta2 \cos^2\frac{k}2 + \sin^2 \frac\theta2 \sin^2\frac{k}2 \right)  = \prod_{k} \frac12 \left(1+ \cos \theta \cos k\right) \\
    && = 2^{1-2n_{\check \tau}} {\cos^{n_{\check \tau}}\theta} \left[ {}_2F_1 \left(n_{\check \tau}, - n_{\check \tau}, \frac12 ; \frac12(1- \sec\theta) \right) +1 \right],
    \label{eq:product}
\eea
where ${}_2F_1(a,b,c;z)$ is the Gaussian hypergeometric function.

We now know the contribution from a cycle $\check\tau$, which can be used to obtain the result of the full solution. Again, the solution in $A$ and $\bar A$ are related by the cyclic permutation $\epsilon$, $\tau_{A} = \epsilon \tau_{\bar A}$. As we discussed before, the maximal number of cycles is $n+1$, i.e., $\check\tau_{(i)}$, $i=1,...,n+1$, where the sum of the length of each cyclic permutation satisfies $\sum_{j=1}^{n+1} n_{\check\tau_{(j)}} = 2n$. Thus we have 
\bea
    && \prod_{x=1}^L \Pf \Big[  \hat S \partial_t  - \Sigma_x \Big] = \left(  \langle \text{EPR} | e^{T \sum_{i=1}^{n+1} H(\check\tau_{(i)})} | \text{EPR} \rangle \right)^{L/2} \\
    && =    \left[ e^{ 2n \sqrt{\Lambda^2 + \mu^2}T} 2^{1-3 n} \cos^{2n}\theta \prod_{i=1}^{n+1} \left[ {}_2F_1 \left(n_{\check\tau_{(i)}}, - n_{\check\tau_{(i)}}, \frac12 ; \frac12(1- \sec\theta) \right) +1 \right]  \right]^{L/2} .
\eea
where in the first line we take a $L/2$ power as there are $L/2$ contributions for the half-chain quasi entropy, i.e., both $A$ and $\bar A$ have $L/2$ number of sites. 

Due to the presence of the measurement, we have to consider the denominator in quasi entropy~(\ref{eq:quasi_disorder}), $\mathbb E \Tr[\tilde \rho^{\otimes n}]$. 
The calculation is paralleled to $\mathbb E \Tr[\tilde \rho^{\otimes n} M_\text{cyc}(A)]$. The difference is that the relation between $\tau_A$ and $\tau_{\bar A}$ should be modified from $\tau_A = \epsilon \tau_{\bar A}$ to $\tau_A = \tau_{\bar A}$, and as we discussed around~(\ref{eq:denominator}) in the previous section, to maximize the number of cycles, it is given by the identity matrix. 
So taking $\tau_A$ and $\tau_{\bar A}$ to be the identity which have $2n$ cycles in total, we have gotten the Pfaffian as follows,
\bea
    && \prod_{x=1}^L \Pf \Big[  \hat S \partial_t  - \Sigma_x \Big] = \left(  \langle \text{EPR} | e^{T \sum_{i=1}^{2n} H(1)} | \text{EPR} \rangle \right)^{L/2} \\
    && =    \left( e^{\sqrt{\Lambda^2 + \mu^2}T} 2^{-1} {\cos\theta} \left[ {}_2F_1 \left(1, - 1, \frac12 ; \frac12(1- \sec\theta) \right) +1\right] \right)^{n L} \\
    && = e^{n\sqrt{\Lambda^2 + \mu^2}TL} \left( \cos \frac{\theta}2 \right)^{2nL}.
\eea
where $H(1)$ in the first line denotes~(\ref{eq:cycle}) with trivial cycle $\check\tau = 1$.  
On the other hand, the calculation of~(\ref{eq:onshell_unitary}),~(\ref{eq:onshell_monitor}) and~(\ref{eq:decay2}) is the same as before. Thus, all the time dependent contributions cancel when we take the ratio between these two quantities, namely, $\frac{\mathbb E \Tr[\tilde \rho^{\otimes n}S_\text{cyc}(A)]}{\mathbb E \Tr[\tilde \rho^{\otimes n}]}$. And finally, we get the quasi entropy, 
\bea \label{eq:quasi}
    e^{(1-n) S_A^{(n)}}= 2^{L(n-1)} \sum_{\mu=1}^{C_n} \left[ \frac{2^{1-3 n} \cos^{2n}\theta \prod_{i=1}^{n+1} \left[  {}_2F_1 \left(n_{\check\tau^{\mu}_{(i)}}, - n_{\check\tau^{\mu}_{(i)}}, \frac12 ; \frac12(1- \sec\theta) \right) +1\right]}{\cos^{4 n}\frac\theta2} \right]^{NL/2},
\eea
where $\{ \check\tau^{\mu}_{(i)} \}$, $i=1,...,n+1$, is the cycle decomposition of the solution $\tau_A^{\mu}$ and $\tau_{\bar A}^\mu$, and there are $C_n$ number of such solutions, $\mu=1,...,C_n$. $C_n$ is the Catalan number. In the prefactor we also include the degeneracy coming from the local Fermi parity symmetry.

Without measurements, i.e., $\theta=0$, (\ref{eq:quasi}) reduces to two decoupled unitary chains discussed at the end of Section~\ref{sec:cluster}. With the measurement, we are not able to evaluate the product explicitly. It is however clear that as $\theta \rightarrow  \frac{\pi}2 $, the different saddle-point solutions tend to the same form with no correlation between replicas, because $\Lambda = 0$ in~(\ref{eq:solution2}). And according to 
\bea
    \lim_{\theta \rightarrow \frac\pi2} \cos^{n}\theta \left[  {}_2F_1 \left(n_{\check\tau^{\mu}_{(i)}}, - n_{\check\tau^{\mu}_{(i)}}, \frac12 ; \frac12(1- \sec\theta) \right) +1\right] = 2^{n-1} e^{i2\pi n},
\eea
we have $e^{(1-n) S_A^{(n)}}= \left(\frac{2^{1-3n} 2^{n-1} e^{i2\pi n}}{(1/\sqrt2)^{4n}} \right)^{NL/2} = 1$ from~(\ref{eq:quasi}) where all degeneracy disappears as the replica symmetric solution is unique, for any integer $n \ge 2$. This implies that the quasi entropy of any order vanishes at $\theta = \frac\pi2$ in the large-$N$ limit. In the next section, we will consider a special solution, the cyclic symmetric one, and analytically continue the solution to $n \rightarrow 1$ to show the von Neumann entanglement entropy transition.

\subsection{von Neumann entanglement entropy from replica trick}

\begin{figure}
    \centering
    \includegraphics[width=0.5 \textwidth]{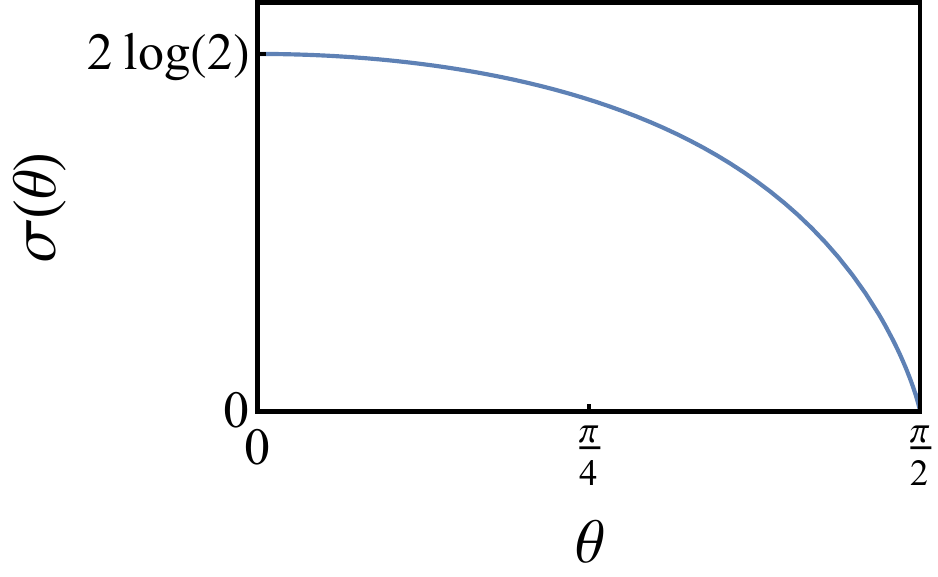}
    \caption{The plot of von Neumann entanglement entropy per site per flavor $\sigma(\theta)$ as a function of the monitoring rate $\theta = \arctan \frac{\mu}{\Lambda}$. }
    \label{fig:line_tension}
\end{figure}

Among the $C_n$ different permutation that satisfy $\tau_A = \epsilon \tau_{\bar A}$ and have the maximal number of cycles, there is a cyclic symmetric one, where $\tau_{\bar A}$ is identity, and $\tau_A$ is itself a cyclic permutation of length $n$~\footnote{In the gravity setup~\cite{penington2019replica}, similar type of solution is considered in taking the $n\rightarrow 1$ limit.}. So the number of cycles is maximal, $n+1$. This is the only nontrivial solution when $n=2$, i.e., $C_2=1$, and we expect it to dominate for $n < 2 $ in the replica limit. Focusing on the contribution of this special solution, we make the following analytic continuation,
\bea
    e^{(1-x) S^{(x)}_A} =  \left[ \frac{ \left( \cos^{2} \frac\theta2\right)^x \left(  \frac{\cos^{x}\theta}{2^{2x-1}} \left[ {}_2F_1 \left(x, - x, \frac12 ; \frac12(1- \sec\theta) \right) +1 \right] \right)}{\cos^{4 x}\frac\theta2} \right]^{NL/2}, 
\eea
where $x$ is a real number~\footnote{We neglect the degeneracy from the Fermi parity transformation since it is $\mathcal{O}(1)$ contribution, and we are interested in $\mathcal{O}(N)$ contribution.}. In the numerator inside the parentheses, the first factor is the contribution from identity $\tau_{\bar A}^{\alpha\beta} = \delta^{\alpha\beta}$, and the second factor is the contribution from the cyclic permutation $\tau_A^{\alpha\beta} = \epsilon^{\alpha\beta}$. We expand the result near $x=1$ to get the von Neumann entanglement entropy
\bea
    S_A = \lim_{x\rightarrow 1} \frac1{1-x} \log \left[ 1 + (1-x) \sigma(\theta) \right]^{NL/2} = \sigma(\theta) \frac{NL}2,
\eea
where the von Neumann entanglement entropy per site per flavor is
\bea
    \sigma(\theta) =  \left( \log 2(1+\sec \theta) + \tan \frac\theta2 \log(\sec\theta - \tan\theta) \right), \quad \tan \theta = \frac{\mu}{\Lambda}.
\eea
A plot of this function is shown in Fig.~\ref{fig:line_tension}. At $\theta = 0$, we recover the maximal von Neumann entanglement entropy $S_A = NL \log 2$. The von Neumann entanglement entropy density vanishes smoothly at $\theta = \frac\pi2$, where the von Neumann entanglement entropy becomes area-law. It shows clearly that the measurement-induced phase transition occurs at the unbreaking point of replica symmetry. Near the transition, we can expand the von Neumann entanglement entropy to get
\bea
    S_A = \frac{NL}2 \left[\log\left( \frac\pi2 - \theta \right) - \log 2e \right] \left( \frac\pi2 - \theta \right).
\eea
Thus, by noticing the relation between $\theta$ and $\Lambda$ in~(\ref{eq:lambda}), the entanglement density vanishes as $(1- \frac{\mu}J)^{1/2} \log( 1- \frac{\mu}J)$ near the critical point $\mu = J$. This result differs from the known universality classes. 

It is instructive to consider a more general case for $L_A \ne L_{\bar A}$. When $L_A < L_{\bar A}$, the solution is given by the identity in $L_{\bar A}$ and the replica symmetric permutation matrix in $L_A$, because the identity is onshell in the bulk and satisfies the boundary condition (see Appendix~\ref{append:saddle}). Then the von Neumann entanglement entropy for this solution is
\bea
    && e^{(1-n) S^{(n)}_A} = 
    \frac{(\cos^{2n} \frac{\theta}2)^{N L_{\bar A}} \left(  \frac{\cos^{n}\theta}{2^{2n-1}} \left[ {}_2F_1 \left(n, - n, \frac12 ; \frac12(1- \sec\theta) \right) +1 \right]  \right)^{N L_A}}{(\cos^{2n} \frac\theta2)^{N(L_A + L_{\bar A})}}.
\eea
It is clear that the numerator has two factors, the first one from $L_{\bar A}$ and the second one from $L_{\bar A}$. Taking a similar analytic continuation, we get the same result, $S_A = \sigma(\theta) N L_A $.

\section{Conclusion and discussion} \label{sec:conclusion}

To conclude, we introduced a large-$N$ model in which the von Neumann entropy can be calculated analytically. We obtained the R\'enyi entropy by summing over a class of solutions related to the permutation group for the non-measured case, and we analytically established a von Neumann entanglement entropy transition using the replica trick in the meansured case. The result in Section~\ref{sec:cluster} strongly suggests that the late-time von Neumann entanglement entropy of a chaotic unitary system is dominated by saddle points that connect different replicas. Interestingly, the degeneracy of these replica non-diagonal solutions play an important role in determining the entanglement entropy, and a single replica non-diagonal solution can lead to an inaccurate density spectrum density of state. For instance, if we neglected the degeneracy given by the Catalan number, we would get $D(\lambda) = 2^{N-2} \delta(\lambda - 2^{-N+2})$. This would imply that all R\'enyi entropies are given by the same value, similar to stabilizer states generated by Clifford circuits. It would be interesting to understand if there is any larger significance to this observation, for example, related to a path integral representation of Clifford dynamics.

The continuous transition between replica symmetry broken and unbroken solutions studied in Section~\ref{sec:monitor} should be contrasted with the discontinuous transition between two similar large-$N$ solutions in the unitary time evolution of R\'enyi entropy~\cite{jian2020note}. As the replica non-diagonal solution is closely related to the replica wormhole observed in the context of black hole information paradox~\cite{penington2019replica, almheiri2019replica}, it is worth speculating about the physics of a monitored black hole. In the current setup, the monitoring is implemented on the full system and causes the restoration of replica symmetry. We expect that by increasing the monitoring of the black hole and its environment at late time, the replica wormhole continuously disappears. Because the so-called entanglement island is obtained by continuing the cyclic symmetric wormhole~\cite{penington2019replica}, it should also disappear with sufficiently frequent measurements. An interesting question is to monitor the environment, and study the effect of such measurements on the black hole. Though the measurement destroys the entanglement between black hole and radiation, the state remaining in black hole is expected to be maximally scrambled. We leave this exploration as a future work.

More broadly, the approach of mapping the path integral of multi-replicas to the transition amplitude of quantum states provides a general tool in evaluations of entropy related quantities. We expect generalization of such an approach to other models that admit saddle-point analysis to be straightforward. For example, it is of value to generalize and utilize this tool in bosonic models~\cite{bentsen2021measurement}. In this case, the emergent transition amplitude will governed by a spin model living in the replica space. The transition amplitude for a general $n$ might be complicated, but we expect that a similar cyclic symmetric replica non-diagonal solution dominates near $n \approx 1$, which could be analysed in detail.

Finally, the model defined by the action~(\ref{eq:action2}) is distinct from the model we considered previously in Ref.~\cite{jian2021syk}. The major distinction lies in that the action~(\ref{eq:LRhamiltonian}) conserves local Fermi parity, whereas the model studied in Ref.~\cite{jian2021syk} breaks the local Fermi parity by two Majorana hopping between sites. If we consider $n=2$, the Landau-Ginzbug action for $q=4$ would be 
\bea \label{eq:Z4_model}
 \frac{I_{\text{eff}}}{N} &=& \int dt dx \Big( \frac12 \big[ (\partial_t \vec \phi)^2 + \phi_1^2 (\partial_x \phi_1)^2 + \phi_2^2 (\partial_x \phi_2)^2 \big]  + r \vec \phi^2 + \lambda |\vec \phi|^4 + \lambda' ( \phi_1^4 + \phi_2^4) \Big),
\eea
where $\vec \phi = (\phi_1, \phi_2) $ is a two-dimensional order parameter related to rotations between two replicas. The space derivative appears in the forth order term, and it leads to a distinct universality class. It will be interesting to derive the Landau-Ginzburg action in the limit $n \rightarrow 1$.

\section*{Acknowledgements}

We would like to thank Pengfei Zhang and Zhuo-Yu Xian for useful discussions. This work is supported by the Simons Foundation via the It From Qubit Collaboration. The work of BGS is also supported in part by the AFOSR under grant number FA9550-19-1-0360.

\appendix

\section{Permutation operator for fermionic state} \label{append:permutation}

Consider a Hilbert space consisting of $N$ Majorana fermions, $\psi_1, ..., \psi_N$, $\{\psi_i, \psi_j \} = \delta_{ij}$, and $N$ is an even integer. It is convenient to double the Hilbert space and work in a state language. To do that, we introduce another set of Majoran fermion called $\chi_1, ..., \chi_N$, $\{\chi_i, \chi_j \} = \delta_{ij}$, and the maximally entangled state between these two sets of Majrana fermions, $|\text{EPR}\rangle$, defined by
\bea
    (\psi_j + i \chi_j) | \text{EPR} \rangle=0, \quad \forall j= 1,...,N.
\eea
Then any operator acting on the $\psi$ Hilbert space can be uniquely mapped to a state in the doubled Hilbert space by acting it on the EPR state $O_\psi \rightarrow O_\psi | \text{EPR} \rangle$. 

Let us first consider a simpler case with $N=2$. By pairing the two Majorana into a single complex fermion, $c_\psi= \frac{\psi_1+i\psi_2}{\sqrt2}$, $c_\psi^\dag = \frac{\psi_1-i \psi_2}{\sqrt{2}}$ and $c_\chi= i \frac{\chi_1-i\chi_2}{\sqrt2}$, $c_\chi^\dag = -i \frac{\chi_1+i \chi_2}{\sqrt{2}}$ the Hilbert space is spanned by
\bea
    \{ |00\rangle, |10\rangle= c_\psi^\dag |00\rangle, |01\rangle = c_\chi^\dag |00\rangle , |11\rangle = c_\psi^\dag c_\chi^\dag |00\rangle \},
\eea
with the EPR state expressed as $|\text{EPR} \rangle = \frac{|00\rangle + |11\rangle}{\sqrt2}$. It is convenient to label it as $|\text{EPR} \rangle = \frac{|0\bar 0\rangle + |1\bar 1\rangle}{\sqrt2} = \sum_{a} \frac{|a\bar a\rangle}{\sqrt2}$. 
For arbitrary even integer $N$, we have the similar expression $|\text{EPR} \rangle = \sum_{a} \frac{|a\bar a\rangle}{2^{N/2}}$. 

We take $n$ copies of the doubled Hilbert space, and denote the Majorana operators as $\psi_i^\alpha, \chi_i^\alpha$, $i=1,2,...,N$, $\alpha = 1,2, ..., n$. The cyclic permutation operator is defined to take
\bea \label{eq:define_cyc}
    M_\text{cyc} |a_1 \bar a_1 a_2 \bar a_2 a_3 \bar a_3 ... a_n \bar a_n \rangle \rightarrow  |a_1 \bar a_n a_2 \bar a_1 a_3 \bar a_2 ... a_n \bar a_{n-1} \rangle ,
\eea
and it is trivial to show that 
\bea
    \Tr[\rho^n] = \langle \text{EPR}|^{\otimes n} \rho^{\otimes n} M_\text{cyc} | \text{EPR} \rangle^{\otimes n}, \quad | \text{EPR} \rangle^{\otimes n} = 2^{-nN/2} \sum_{a_1,...,a_n} |a_1 \bar a_1 a_2 \bar a_2 a_3 \bar a_3 ... a_n \bar a_n \rangle. \nn \\
\eea
It is also easy to show that the following operator gives~(\ref{eq:define_cyc}),
\bea
     M_\text{cyc} = \prod_{i=1}^N e^{\frac\pi2 \chi^n_i \chi^{n-1}_i} e^{\frac\pi2 \chi^{n-1}_i \chi^{n-2}_i} ... e^{\frac\pi2 \chi^3_i \chi^{2}_i} e^{\frac\pi2 \chi^2_i \chi^{1}_i}.
\eea
In terms of the operator in the $\psi$ Hilbert space, the cyclic permutation becomes
\bea
    M_\text{cyc} = \prod_{i=1}^N e^{\frac\pi2 \psi^1_i \psi^2_i} e^{\frac\pi2 \psi^2_i \psi^3_i} ... e^{\frac\pi2 \psi^{n-2}_i \psi^{n-1}_i} e^{\frac\pi2 \psi^{n-1}_i \psi^{n}_i},
\eea
with the action on the $\psi_i^\alpha$ field
\bea \label{eq:permutation}
    M_\text{cyc} \psi_i^{\alpha} M_\text{cyc}^\dag = \sum_\beta \sgn(\alpha-\beta) \delta^{\alpha+1,\beta} \psi_i^\beta, \quad \forall i =1,...,N,
\eea
where we define $\delta^{n+1,\beta} = \delta^{1,\beta}$ when $\alpha = n$. Thus, (\ref{eq:permutation}) is the cyclic permutation operator for fermionic states, and is used in Section~\ref{sec:cluster}.

\section{Saddle-point solutions} \label{append:saddle}

The equations of motion and boundary conditions for the saddle-point analysis given in the following for general $n$ is a complicated dynamical problem,
\bea 
    \partial_t \hat S \hat G_x(t,t) \hat S &=& [\hat \Sigma_x(t), \hat G_x(t,t)], \\
    \Sigma_{x,ss'}^{\alpha\beta}(t) &=& c_{ss'}\Big[ J (2G_{x,ss'}^{\alpha\beta}) +  U (2G_{x,ss'}^{\alpha\beta})^{q/2-1} [M_{(x)}]^{\alpha\gamma}_s [M_{(x)}]^{\beta\delta}_{s'} (2G_{\bar x,ss'}^{\gamma\delta})^{q/2} \Big], \\
    \psi_{x,+}^\alpha(0) &=& - \psi^\alpha_{x,-}(0), \quad \psi_{x,+}^\alpha (T) = \psi_{x,-}^\alpha(T), \quad \forall \alpha =1,...,n.
\eea
One can show that there is a conserved quantity for each site, given by
\bea
    \Tr(\hat S \hat G_x \hat S \hat G_x) = \text{const},
\eea
where the constant is set by the boundary condition. This follows from
\bea
    \frac{d}{dt} \Tr(\hat S \hat G_x \hat S \hat G_x) = 2 \Tr\left( \frac{d(\hat S \hat G_x \hat S)}{dt} \hat G_x \right) = 2 \Tr\left( [\Sigma_x, G_x] \hat G_x \right) = 0.
\eea
Except for this conserved quantity, in general, the problem becomes a coupled nonlinear differential equation that has no general solution. For $n=2$, we explicitly write down the equations of motion
\bea
    \frac{dx_1}{dt} &=& 4 J x_2 z_1 + 2 U x_2 z_1 (y_1^2 + w^2) , \\
    \frac{dx_2}{dt} &=& -4 J x_1 z_1 - 2 U x_1 z_1 (y_2^2 + w^2), \\
    \frac{dz_1}{dt} &=& 2U x_1 x_2 (y_1^2 - y_2^2), \\
    \frac{dy_1}{dt} &=& 4 J y_2 w_1 + 2U y_2 w_1 (x_1^2 + z^2 ), \\
    \frac{dy_2}{dt} &=& -4J y_1 w_1 - 2 U y_1 w_1 (x_2^2 + z^2), \\
    \frac{dw_1}{dt} &=& 2 U y_1 y_2 (x_1^2 - x_2^2 ),
\eea
where the coefficients are defined via
\bea
    G_A(t,t) =  \frac12\left( \sum_{k=1}^n x_k(t) X_k + \sum_{m=1}^{[\frac{n}2]} z_m(t) Z_m \right), \quad G_{\bar A}(t,t) = \frac12 \left( \sum_{k=1}^n y_k(t) X_k + \sum_{m=1}^{[\frac{n}2]} w_m(t) Z_m \right), \nn \\
\eea
with 
\bea \label{eq:basis}
    X_k = \left( \ba{cccc} 0 & - \sigma^{-k} \\ \sigma^k & 0 \ea \right), && \quad k = 1, ..., n, \\
    Z_m =  \frac12 \left( \ba{cccc} \sigma^m - \sigma^{-m} & 0  \\ 0 & \sigma^{-m} - \sigma^m \ea \right), && \quad m = 1,..., \left[\frac{n}{2} \right]. 
\eea
Here $\left[ \frac{n}2\right]$ is the largest integer that is less than $\frac{n}2$, and $\sigma^{\alpha\beta} = \sgn(\alpha-\beta) \delta^{\alpha+1,\beta}$ is the permutation matrix with a proper sign to be consistent with the even Fermi parity. When $\alpha = n$, the symbol means $\delta^{n+1,\beta} = \delta^{1,\beta}$. 

Although we have only explicitly write down the equation for $n=2$, it can be seen from the definition of the matrices that these type of coefficients form a closed set of dynamical variables. It reduces the number of variable from $\sim n^2$ to $\sim n$. For $n=2$, this representation actually captures all independent variables. In terms of this set of variables, the conserved quantity is given by
\bea \label{eq:conserved}
    \sum_{k=1}^{n} x_k^2 - \sum_{m=1}^{[\frac{n}2]} z_m^2 = \text{const}, \quad \sum_{k=1}^{n} y_k^2 - \sum_{m=1}^{[\frac{n}2]} w_m^2 = \text{const}.
\eea

To further simplify the problem, we assume that in $\bar A$ the conserved quantity is given by $y_1^2+y_2^2 - w_1^2=1$ (as we discuss in the following, this is required to have a time-independent action), and since the boundary condition is $y_2=-1$, it uniquely determines the steady solution $y_1 = w_1 = 0$ and $y_2 = -1$. The equations of motion reduce to
\bea
    \frac{dx_1}{dt} &=& 4 J x_2 z_1  , \\
    \frac{dx_2}{dt} &=& -(4 J + 2U) x_1 z_1 , \\
    \frac{dz_1}{dt} &=& -2U x_1 x_2 , 
\eea
and $x_1^2 + x_2^2 - z_1^2 $ is conserved. There is a discrete symmetry given by multiplying any two of the three variables $x_1, x_2, z_1$ by a minus one, leading to four-fold degenerate nontrivial solutions. Among these solutions, half of them are determined by the initial value of $x_2$, i.e., by the boundary condition, and given the boundary condition, there are still two solutions related by the Fermi parity transformation discussed in the main text. In the following, we will discuss one explicit solution.

These differential equations can be solved via Jacobi elliptic functions, 
\bea
    x_1 &=& \sqrt{c_1} \sn \left(2\sqrt{U(2J+U) c_2} (t-t_0), \frac{c_1}{c_2} \right), \\
    x_2 &=& - \sqrt{\frac{(2J+U)c_1}{2J}} \left(1- \sn\left( 2\sqrt{U(2J+U) c_2} (t-t_0), \frac{c_1}{c_2} \right) \right)^{1/2}, \\
    z_1 &=& - \sqrt{\frac{Uc_2}{2J}} \dn \left( 2\sqrt{U(2J+U) c_2} (t-t_0), \frac{c_1}{c_2} \right),
\eea
where $\sn(u,c)$ and $\dn(u,c)$ are the Jacobi elliptic functions. $c_1, c_2, t_0$ are integral constants to be determined below. $t_0$ is the shift of time argument, and $c_1, c_2$ are related to the conserved quantity by $x_1^2 + x_2^2 - z_1^2 = c_1 + \frac{U}{2J} (c_1-c_2) $. 

\begin{figure}
    \centering
    \includegraphics[width=0.45\textwidth]{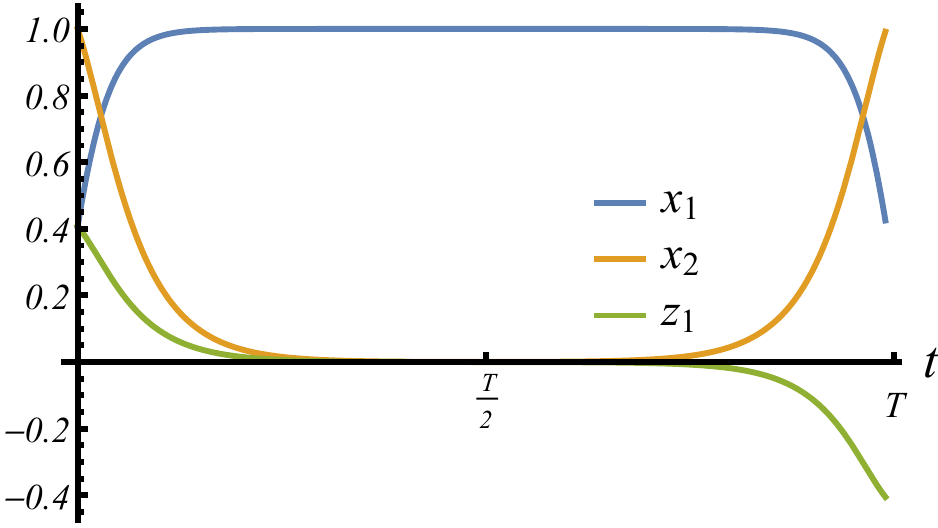}
    \caption{A plot of the solution of $x_1, x_2, z_1$. We choose $J=1$, $U=0.4$, $c_1=1$, $c_2=1-10^{-7}$.}
    \label{fig:solution}
\end{figure}

The conserved quantity is related to the time dependence of the onshell action. Because the action is dimensionless, it can only depend on $JT$ and $UT$. Taking the derivative with respect to $T$, we have
\bea
    \frac{d I(J T, U T)}{d T} &=& \frac{J}{T} \frac{\partial I}{\partial J} + \frac{U}{T} \frac{\partial I}{\partial U} \nn \\
    &=& -c_{ss'} \int dt  \Big[ \frac{J}{8T}   \sum_x (2G_{x,ss'}^{\alpha\beta}(t,t))^2  + \frac{U}{8T} (2G_{1,ss'}^{\alpha\beta}(t,t))^{2} M^{\alpha\gamma}_s M^{\beta\delta}_{s'} (2G_{2,ss'}^{\gamma\delta}(t,t))^{2} \Big] \nn \\
    &=& - \int dt  \Big[ \frac{J}{2T}   (x_1(t)^2 + x_2(t)^2 - z_1(t)^2 - 1) + \frac{U}{2T} (x_1(t)^2-1) \Big],
\eea
where we have plugged in $y_1 = w_1 = 0$ and $y_2 = -1$. In the third line, the first term in parentheses is given by the conserved quantity. To have a time-independent result, the conserved quantity should be $x_1(t)^2 + x_2(t)^2 - z_1(t)^2 =1$. The second term vanishes at long times when $x_1(t) \rightarrow 1$. With these two facts, $c_1$, $c_2$ can be determined to be $c_1 = 1$ and $c_2 = 1$ as $T \rightarrow \infty$. The boundary condition at $t=0$ fixes $t_0 = \frac{1}{4\sqrt{U(2J+U)}} \cosh^{-1} \frac{J+U}J$ such that $x_2(0) = -1$. An example is shown in Fig.~\ref{fig:solution}, where we allow $c_2$ to be slightly less than $1$ to capture the boundary condition at $t=T$. As $c_2$ approaches $1$, the time $T$ approaches $\infty$. In the long-time limit, the solution becomes,
\bea
    x_1 &=&  \tanh \left( 2\sqrt{U(2J+U)} (t-t_0) \right), \\
    x_2 &=& -\sqrt{\frac{2J+U}{2J}}  \frac1{\cosh \left( 2\sqrt{U(2J+U)} (t-t_0) \right) }, \\
    z_1 &=& -\sqrt{\frac{U}{2J}} \frac1{\cosh \left( 2\sqrt{U(2J+U)} (t-t_0) \right) },
\eea
where indeed the solution in the bulk takes the form of~(\ref{eq:solution}), and the effect of the boundary conditions decreases exponentially fast into the bulk. Now in the long-time limit, the onshell action tends to a constant determined by the ratio $U/J$ because it is the only dimensionless quantity when $T \rightarrow \infty$. The long-time onshell action cannot depend on $U(2J+U)$, so we can take this advantage to send $U(2J+U) \rightarrow \infty$ without changing the onshell action. In this case, the solution is given by~(\ref{eq:solution}). Therefore, although~(\ref{eq:solution}) does not satisfy the boundary condition in any finite $U(2J+U)$, the action evaluated via~(\ref{eq:solution}) gives the correct answer in the long-time limit. The argument implicitly states
\bea \label{eq:limit}
    \lim_{E \rightarrow \infty} I\left[ G_\text{exact}, \Sigma_\text{exact} \right] =  I\left[\lim_{E  \rightarrow \infty}  G_\text{exact}, \lim_{E \rightarrow \infty} \Sigma_\text{exact} \right] =  I\left[ G, \Sigma \right],
\eea
where $E$ is a microscopic energy scale in the model, i.e., $U= uE$, $J= j E$~\footnote{This is same as taking $T \rightarrow \infty$, because the action is dimensionless and depends on $TE$ only.}. In the above limit, $U/J = u/j$ is kept fixed. 
$G_\text{exact}$ and $\Sigma_\text{exact}$ denote the exact saddle-point solution satisfying the boundary condition, and $G$ and $\Sigma$ in the last equality is the solution given in~(\ref{eq:solution}). 
This must be true because the long-time limit of the R\'enyi entropy and von Neumann entanglement entropy of the Brownian SYK clusters saturate to a constant set by the dimension of Hilbert space and the symmetry of the model independent of any microscopic energy scale~\cite{page1993average,jian2021quantum,stanford2021subleading}. And the evaluation of the von Neumann entanglement entropy from the permutation solution~(\ref{eq:solution}) correctly gives the Page value (see Section~\ref{sec:cluster}).

For general $n$, we expect the same situation happens. The conserved quantity~(\ref{eq:conserved}) is given by $1$, and the bulk is dominated by one of the permutation matrix $X_k$ given by~(\ref{eq:basis}) to have a time-independent onshell action. 
When the measurement is turned on, the addition of~(\ref{eq:monitor}) is an integration over the Green's function, for which we expect~(\ref{eq:limit}) still holds. Thus, in Section~\ref{sec:monitor}, we consider a generalization of solutions of the form~(\ref{eq:solution}) to the measurement case, c.f.~(\ref{eq:solution2}), and study how the measurement affects the solution and causes the entanglement transition.

\section{Finite products of trigonometric functions} \label{append:product}

In this section we first show that the following identity holds and then use it to derive~(\ref{eq:product}),
\bea \label{eq:identity}
    2(T_n(a) + 1 ) = \begin{cases} 
            \prod_{k=-\frac{n}2+1}^{\frac{n}2} (2a + 2 \cos \frac{(2k+1)\pi}{n}), & n = \text{even}, \\
            \\
            \prod_{k=- \frac{n-1}2}^{\frac{n-1}2} (2a + 2 \cos \frac{2k\pi}{n}), & n = \text{odd}.
            \end{cases}
\eea
for $k$ being integers, where $T_n(a)$ is the Chebyshev polynomial of the first kind, namely, $T_n(\cos \phi) = \cos n \phi$. 
For an even integer $n$, note that~(\ref{eq:identity}) is equivalent to
\bea \label{eq:even}
    x^n + x^{-n} + 2 \cos n \varphi = \prod_{k=-\frac{n}2+1}^{\frac{n}2} \left[ x + x^{-1} - 2\cos\left(\varphi + \frac{(2k-1)\pi}n \right) \right],
\eea
by setting $\varphi = \pi $ and $x = e^{i \phi}$. The left-hand side of~(\ref{eq:even}) has the roots,
\bea
    x = e^{i(\varphi + \frac{(2k-1)\pi}n)}, \quad k \in \{ -\frac{n}2+1,-\frac{n}2+2,..., \frac{n}2 \}.
\eea
By noticing 
\bea
    x + x^{-1} = 2 \cos \left( \varphi + \frac{(2k-1)\pi}n \right),
\eea
we arrive the right-hand side of~(\ref{eq:even}). The proof for an odd integer $n$ is similar. (\ref{eq:identity}) is equivalent to 
\bea \label{eq:odd}
    x^n + x^{-n} - 2 \cos n \varphi = \prod_{k=-\frac{n-1}2}^{-\frac{n-1}2} \left[ x + x^{-1} - 2\cos\left(\varphi + \frac{2k\pi}n \right) \right],
\eea
by setting $\varphi = \pi $ and $x = e^{i \phi}$. The left-hand side of~(\ref{eq:odd}) has the roots,
\bea
    x = e^{i(\varphi + \frac{2k\pi}n)}, \quad k \in \{ -\frac{n-1}2,-\frac{n-1}2+1,..., \frac{n-1}2 \}.
\eea
Then by noticing 
\bea
    x + x^{-1} = 2 \cos \left( \varphi + \frac{2k\pi}n \right),
\eea
we get the right-hand side of~(\ref{eq:even}).

Using the identity~(\ref{eq:identity}), we can derive~(\ref{eq:product}). 
The derivation is simply given here. For even integers $n$,
\bea
    &&\prod_{k=-\frac{n}2+1}^{\frac{n}2} \frac12 \left(1+ \cos \theta \cos k\right)  = 2^{-2n} \cos^n\theta \prod_{k=-\frac{n}2+1}^{\frac{n}2} \left( \frac2{\cos \theta}+ 2 \cos k\right) \\
    && = 2^{1-2n} \cos^n\theta(T_n(\sec\theta) + 1)
    = 2^{1-2n} {\cos^{n}\theta} \left[  {}_2F_1 \left(n, -n, \frac12 ; \frac12(1- \sec\theta) \right) +1 \right],
\eea
and for odd integers $n$,
\bea
    &&\prod_{k=-\frac{n-1}2}^{\frac{n-1}2} \frac12 \left(1+ \cos \theta \cos k\right)  = 2^{-2n} \cos^n\theta \prod_{k=-\frac{n-1}2}^{\frac{n-1}2} \left( \frac2{\cos \theta}+ 2 \cos k\right) \\
    && = 2^{1-2n} \cos^n\theta(T_n(\sec\theta) + 1)
    = 2^{1-2n} {\cos^{n}\theta} \left[ {}_2F_1 \left(n, - n, \frac12 ; \frac12(1- \sec\theta) \right) +1 \right].
\eea
where in the last step of the derivation, we use the fact that the Chebyshev polynomial is equal to the Gaussian hypergeometic function, 
\bea
    T_x(\cos\phi) = {}_2F_1 \left(x, - x, \frac12 ; \frac12(1- \cos\phi) \right), 
\eea
for integer $x$. 
Then we use the fact that the Gaussian hypergeometric function is defined for real number $x$ to make analytical continuations.

\bibliographystyle{jhep}
\bibliography{reference}

\providecommand{\href}[2]{#2}\begingroup\raggedright\begin{thebibliography}{10}

\bibitem{li2018quantum}
Y.~Li, X.~Chen and M.~P. Fisher, \emph{Quantum zeno effect and the many-body
  entanglement transition}, {\emph{Physical Review B} {\bf 98} (2018) 205136}.

\bibitem{li2019measurement}
Y.~Li, X.~Chen and M.~P. Fisher, \emph{Measurement-driven entanglement
  transition in hybrid quantum circuits}, {\emph{Physical Review B} {\bf 100}
  (2019) 134306}.

\bibitem{skinner2019measurement}
B.~Skinner, J.~Ruhman and A.~Nahum, \emph{Measurement-induced phase transitions
  in the dynamics of entanglement}, {\emph{Physical Review X} {\bf 9} (2019)
  031009}.

\bibitem{gullans2020dynamical}
M.~J. Gullans and D.~A. Huse, \emph{Dynamical purification phase transition
  induced by quantum measurements}, {\emph{Physical Review X} {\bf 10} (2020)
  041020}.

\bibitem{chan2019unitary}
A.~Chan, R.~M. Nandkishore, M.~Pretko and G.~Smith, \emph{Unitary-projective
  entanglement dynamics}, {\emph{Physical Review B} {\bf 99} (2019) 224307}.

\bibitem{vasseur2019entanglement}
R.~Vasseur, A.~C. Potter, Y.-Z. You and A.~W. Ludwig, \emph{Entanglement
  transitions from holographic random tensor networks}, {\emph{Physical Review
  B} {\bf 100} (2019) 134203}.

\bibitem{choi2020quantum}
S.~Choi, Y.~Bao, X.-L. Qi and E.~Altman, \emph{Quantum error correction in
  scrambling dynamics and measurement-induced phase transition},
  {\emph{Physical Review Letters} {\bf 125} (2020) 030505}.

\bibitem{jian2020measurement}
C.-M. Jian, Y.-Z. You, R.~Vasseur and A.~W. Ludwig, \emph{Measurement-induced
  criticality in random quantum circuits}, {\emph{Physical Review B} {\bf 101}
  (2020) 104302}.

\bibitem{bao2020theory}
Y.~Bao, S.~Choi and E.~Altman, \emph{Theory of the phase transition in random
  unitary circuits with measurements}, {\emph{Physical Review B} {\bf 101}
  (2020) 104301}.

\bibitem{zabalo2020critical}
A.~Zabalo, M.~J. Gullans, J.~H. Wilson, S.~Gopalakrishnan, D.~A. Huse and
  J.~Pixley, \emph{Critical properties of the measurement-induced transition in
  random quantum circuits}, {\emph{Physical Review B} {\bf 101} (2020) 060301}.

\bibitem{gullans2020scalable}
M.~J. Gullans and D.~A. Huse, \emph{Scalable probes of measurement-induced
  criticality}, {\emph{Physical review letters} {\bf 125} (2020) 070606}.

\bibitem{li2020conformal}
Y.~Li, X.~Chen, A.~W. Ludwig and M.~Fisher, \emph{Conformal invariance and
  quantum non-locality in hybrid quantum circuits}, {\emph{arXiv preprint
  arXiv:2003.12721} (2020) }.

\bibitem{iaconis2020measurement}
J.~Iaconis, A.~Lucas and X.~Chen, \emph{Measurement-induced phase transitions
  in quantum automaton circuits}, {\emph{Physical Review B} {\bf 102} (2020)
  224311}.

\bibitem{nahum2021measurement}
A.~Nahum, S.~Roy, B.~Skinner and J.~Ruhman, \emph{Measurement and entanglement
  phase transitions in all-to-all quantum circuits, on quantum trees, and in
  landau-ginsburg theory}, {\emph{PRX Quantum} {\bf 2} (2021) 010352}.

\bibitem{li2021statistical}
Y.~Li and M.~P. Fisher, \emph{Statistical mechanics of quantum error correcting
  codes}, {\emph{Physical Review B} {\bf 103} (2021) 104306}.

\bibitem{sang2021measurement}
S.~Sang and T.~H. Hsieh, \emph{Measurement-protected quantum phases},
  {\emph{Physical Review Research} {\bf 3} (2021) 023200}.

\bibitem{lavasani2021measurement}
A.~Lavasani, Y.~Alavirad and M.~Barkeshli, \emph{Measurement-induced
  topological entanglement transitions in symmetric random quantum circuits},
  {\emph{Nature Physics} {\bf 17} (2021) 342--347}.

\bibitem{ippoliti2021entanglement}
M.~Ippoliti, M.~J. Gullans, S.~Gopalakrishnan, D.~A. Huse and V.~Khemani,
  \emph{Entanglement phase transitions in measurement-only dynamics},
  {\emph{Physical Review X} {\bf 11} (2021) 011030}.

\bibitem{bao2021symmetry}
Y.~Bao, S.~Choi and E.~Altman, \emph{Symmetry enriched phases of quantum
  circuits}, {\emph{arXiv preprint arXiv:2102.09164} (2021) }.

\bibitem{bentsen2021measurement}
G.~Bentsen, S.~Sahu and B.~Swingle, \emph{Measurement-induced purification in
  large-n hybrid brownian circuits}, {\emph{arXiv preprint arXiv:2104.07688}
  (2021) }.

\bibitem{jian2021syk}
S.-K. Jian, C.~Liu, X.~Chen, B.~Swingle and P.~Zhang, \emph{Syk meets
  non-hermiticity ii: measurement-induced phase transition}, {\emph{arXiv
  preprint arXiv:2104.08270} (2021) }.

\bibitem{li2021entanglement}
Y.~Li, S.~Vijay and M.~Fisher, \emph{Entanglement domain walls in monitored
  quantum circuits and the directed polymer in a random environment},
  {\emph{arXiv preprint arXiv:2105.13352} (2021) }.

\bibitem{jian2021quantum}
S.-K. Jian, C.~Liu, X.~Chen, B.~Swingle and P.~Zhang, \emph{Quantum error as an
  emergent magnetic field}, {\emph{arXiv preprint arXiv:2106.09635} (2021) }.

\bibitem{yang2021entanglement}
Z.-C. Yang, Y.~Li, M.~Fisher and X.~Chen, \emph{Entanglement phase transitions
  in random stabilizer tensor networks}, {\emph{arXiv preprint
  arXiv:2107.12376} (2021) }.

\bibitem{kitaev2015simple}
A.~Kitaev, \emph{A simple model of quantum holography}, {\emph{talk given at
  the KITP Program: entanglement in strongly-correlated quantum matter} (2015)
  }.

\bibitem{sachdev1993gapless}
S.~Sachdev and J.~Ye, \emph{Gapless spin-fluid ground state in a random quantum
  heisenberg magnet}, {\emph{Physical review letters} {\bf 70} (1993) 3339}.

\bibitem{maldacena2016remarks}
J.~Maldacena and D.~Stanford, \emph{Remarks on the sachdev-ye-kitaev model},
  {\emph{Physical Review D} {\bf 94} (2016) 106002}.

\bibitem{saad2018semiclassical}
P.~Saad, S.~H. Shenker and D.~Stanford, \emph{A semiclassical ramp in syk and
  in gravity}, {\emph{arXiv preprint arXiv:1806.06840} (2018) }.

\bibitem{sunderhauf2019quantum}
C.~S{\"u}nderhauf, L.~Piroli, X.-L. Qi, N.~Schuch and J.~I. Cirac,
  \emph{Quantum chaos in the brownian syk model with large finite n: Otocs and
  tripartite information}, {\emph{Journal of High Energy Physics} {\bf 2019}
  (2019) 1--44}.

\bibitem{liu2021non}
C.~Liu, P.~Zhang and X.~Chen, \emph{Non-unitary dynamics of sachdev-ye-kitaev
  chain}, {\emph{SciPost Physics} {\bf 10} (2021) Art--No}.

\bibitem{jian2020note}
S.-K. Jian and B.~Swingle, \emph{Note on entropy dynamics in the brownian syk
  model}, {\emph{arXiv preprint arXiv:2011.08158} (2020) }.

\bibitem{stanford2021subleading}
D.~Stanford, Z.~Yang and S.~Yao, \emph{Subleading weingartens}, {\emph{arXiv
  preprint arXiv:2107.10252} (2021) }.

\bibitem{page1993average}
D.~N. Page, \emph{Average entropy of a subsystem}, {\emph{Physical review
  letters} {\bf 71} (1993) 1291}.

\bibitem{napp2019efficient}
J.~Napp, R.~L. La~Placa, A.~M. Dalzell, F.~G. Brandao and A.~W. Harrow,
  \emph{Efficient classical simulation of random shallow 2d quantum circuits},
  {\emph{arXiv preprint arXiv:2001.00021} (2019) }.

\bibitem{kitaev2001unpaired}
A.~Y. Kitaev, \emph{Unpaired majorana fermions in quantum wires},
  {\emph{Physics-uspekhi} {\bf 44} (2001) 131}.

\bibitem{wiseman1996quantum}
H.~Wiseman, \emph{Quantum trajectories and quantum measurement theory},
  {\emph{Quantum and Semiclassical Optics: Journal of the European Optical
  Society Part B} {\bf 8} (1996) 205}.

\bibitem{penington2019replica}
G.~Penington, S.~H. Shenker, D.~Stanford and Z.~Yang, \emph{Replica wormholes
  and the black hole interior}, {\emph{arXiv preprint arXiv:1911.11977} (2019)
  }.

\bibitem{gu2017spread}
Y.~Gu, A.~Lucas and X.-L. Qi, \emph{Spread of entanglement in a
  sachdev-ye-kitaev chain}, {\emph{Journal of High Energy Physics} {\bf 2017}
  (2017) 1--44}.

\bibitem{chen2020replica}
Y.~Chen, X.-L. Qi and P.~Zhang, \emph{Replica wormhole and information
  retrieval in the syk model coupled to majorana chains}, {\emph{Journal of
  High Energy Physics} {\bf 2020} (2020) 1--26}.

\bibitem{altland2010condensed}
A.~Altland and B.~D. Simons, \emph{Condensed matter field theory}.
\newblock Cambridge university press, 2010.

\bibitem{almheiri2019replica}
A.~Almheiri, T.~Hartman, J.~Maldacena, E.~Shaghoulian and A.~Tajdini,
  \emph{Replica wormholes and the entropy of hawking radiation}, {\emph{arXiv
  preprint arXiv:1911.12333} (2019) }.

\bibitem{jacques1968genre}
A.~Jacques, \emph{Sur le genre d'une paire de substitutions}, {\emph{Comptes
  Rendus de l'Acad{\'e}mie des Sciences, Paris} {\bf 267} (1968) 625--627}.

\bibitem{dulucq1998combinatorial}
S.~Dulucq and R.~Simion, \emph{Combinatorial statistics on alternating
  permutations}, {\emph{Journal of Algebraic Combinatorics} {\bf 8} (1998)
  169--191}.

\end{thebibliography}\endgroup

\end{document}